\documentclass{IEEEtran}

\usepackage{cite}
\usepackage{amsthm, amsmath,amssymb,amsfonts, nccmath}
\usepackage{algorithmic}
\usepackage{graphicx}
\usepackage{subfigure}

\usepackage{hyperref}
\usepackage{color}
\usepackage{times}
\usepackage{url}

\graphicspath{{graphics/}}
\usepackage{multirow}
\usepackage{filecontents, pgfplots}
\usepackage[algoruled]{algorithm2e}
\usepackage{microtype}
\usepackage{tikz}
\usepackage{textcomp}
\usepackage{xcolor}
\usepackage[binary-units=true]{siunitx}
\DeclareSIUnit{\belmilliwatt}{Bm}
\DeclareSIUnit{\dBm}{\deci\belmilliwatt}
\def\BibTeX{{\rm B\kern-.05em{\sc i\kern-.025em b}\kern-.08em
		T\kern-.1667em\lower.7ex\hbox{E}\kern-.125emX}}

\usepackage{xcolor,cite,etoolbox}
\makeatletter 
\pretocmd\@bibitem{\color{black}\csname keycolor#1\endcsname}{}{\fail}
\newcommand\citecolor[1]{\@namedef{keycolor#1}{\color{blue}}}
\makeatother

\makeatletter
\newif\iftag@here

\newcommand*{\taghere}[1][0pt]
{\ifmeasuring@\else
	\global\tag@heretrue
	\tikz[remember picture,overlay]{\coordinate (taghere) at (0pt,#1);}%
	\fi}

\def\place@tag{%
	\iftagsleft@
	\kern-\tagshift@
	\iftag@here
	\global\tag@herefalse
	\tikz[remember picture,overlay]%
	{\path (taghere) -| node[anchor=base]{\rlap{\boxz@}} (0pt,0pt);}%
	\else
	\if1\shift@tag\row@\relax
	\rlap{\vbox{%
			\normalbaselines
			\boxz@
			\vbox to\lineht@{}%
			\raise@tag
	}}%
	\else
	\rlap{\boxz@}%
	\fi
	\kern\displaywidth@
	\fi
	\else
	\kern-\tagshift@
	\iftag@here
	\global\tag@herefalse
	\tikz[remember picture,overlay]%
	{\path  (taghere) -|  node[anchor=base]{\llap{\boxz@}} (0pt,0pt);}%
	\else
	\if1\shift@tag\row@\relax
	\llap{\vtop{%
			\raise@tag
			\normalbaselines
			\setbox\@ne\null
			\dp\@ne\lineht@
			\box\@ne
			\boxz@
	}}%
	\else \llap{\boxz@}%
	\fi
	\fi
	\fi
}
\makeatother

\DeclareMathOperator*{\argmax}{arg\,max}
\DeclareMathOperator*{\maximize}{maximize}

\usepackage[acronym,nomain]{glossaries}
\makeglossaries
\newacronym{swipt}{SWIPT}{simultaneous wireless information and power transfer}
\newacronym{wpt}{WPT}{wireless power transfer}
\newacronym{awgn}{AWGN}{additive white Gaussian noise}
\newacronym{tx}{TX}{transmitter}
\newacronym{ir}{IR}{information receiver}
\newacronym{eh}{EH}{energy harvester}
\newacronym{ap}{AP}{average power}
\newacronym{pp}{PP}{peak power}

\newacronym{rf}{RF}{radio frequency}
\newacronym{dc}{DC}{direct current}
\newacronym{ac}{AC}{alternative current}
\newacronym{papr}{PAPR}{peak-to-average power ratio}
\newacronym{lp}{LPF}{low-pass filter}
\newacronym{rv}{RV}{random variable}
\newacronym{iid}{i.i.d.}{independent and identically distributed}
\newacronym{pdf}{pdf}{probability density function}

\newacronym{dnn}{DNN}{dense neural networks}
\glsunset{dnn}
\newacronym{mdp}{MDP}{Markov decision process}

\newacronym{spr}{LP}{low power}
\newacronym{mpr}{MP}{medium power}
\newacronym{lpr}{HP}{high power}

\makeatletter
\def\ps@IEEEtitlepagestyle{%
	\def\@oddhead{\mycopyrightnotice}%
	\def\@evenhead{}%
}
\def\mycopyrightnotice{%
	{\footnotesize 
		\begin{minipage}{\textwidth}
			\centering
			This article has been accepted for publication in a future issue of this journal. Content may change prior to final publication. \\Citation information: DOI~10.1109/TCOMM.2020.3034359, IEEE Transactions on Communications\hfill
		\end{minipage}
		
	}
	\gdef\mycopyrightnotice{ }
}

\begin{document}

	\newtheorem{proposition}{Proposition}	
	
	\title{Markov Decision Process Based Design of SWIPT Systems: Non-linear EH Circuits, Memory, and Impedance Mismatch}
	
	\author{Nikita~Shanin,~\IEEEmembership{Student Member,~IEEE}, Laura~Cottatellucci,~\IEEEmembership{Member,~IEEE}, and Robert~Schober,~\IEEEmembership{Fellow,~IEEE}}
	
	\maketitle
	
	\begin{abstract}		
		In this paper, we study simultaneous wireless information and power transfer (SWIPT) systems employing practical non-linear energy harvester (EH) circuits. 
		Since the voltage across the reactive elements of realistic EH circuits cannot drop or rise instantaneously, EHs have \emph{memory} which we model with a Markov decision process (MDP). 
		Moreover, since an analytical model that accurately models all non-linear effects and the unavoidable impedance mismatch of EHs is not tractable, we propose a learning based model for the EH circuit. 
		We optimize the input signal distribution for maximization of the harvested power under a constraint on the minimum mutual information between transmitter (TX) and information receiver (IR).
		We distinguish the cases where the MDP state is known and not known at TX and IR. 
		When the MDP state is known, the formulated optimization problem for the harvested power is convex.
		In contrast, if TX and IR do not know the MDP state, the resulting optimization problem is non-convex and solved via alternating optimization, which is shown to yield a limit point of the problem.
		Our simulation results reveal that the rate-power region of the considered SWIPT system depends on the symbol duration, the EH input power level, the EH impedance mismatch, and the type of EH circuit.
		In particular, a shorter symbol duration enables higher bit rates at the expense of a significant decrease in the average harvested power.
		Furthermore, whereas half-wave rectifiers outperform full-wave rectifiers in the low and medium input power regimes, full-wave rectifiers are preferable if the input power at the EH is high\let\thefootnote\relax\footnote{The work was supported in part by German Research Foundation under Project SCHO 831/12-1. This paper was presented in part at the IEEE International Conference on Communications (ICC), Dublin, Ireland, 2020~\cite{Shanin2019}.
		
		The authors are with the Institute of Digital Communications, Friedrich-Alexander-Universit\"{a}t Erlangen-N\"{u}rnberg (FAU), Germany.}.		
	\end{abstract}
\setcounter{footnote}{0} 
	\section{Introduction}
	
	The Internet-of-Things (IoT) and the related tremendous growth of the number of low-power devices have attracted significant attention in recent years. 
	Nevertheless, the problem of efficient recharging or replacement of the batteries of billions of IoT devices, such as wireless sensors and actuators, remains unsolved \cite{Clerckx2019}. 
	A promising exploitable feature to address this problem is the ability of \gls*{rf} signals to transfer not only information but also energy, which can be harvested by these devices.
	This prospect has fueled significant interest in \gls*{swipt} systems \cite{Varshney2008, Zhang2013, Clerckx2019, Pilanawithana2019, Gautam2018, Boshkovska2016, Kang2018, Ma2019, Boshkovska2015, Kang2017, Xiong2017, Kang2020, Zhou2013, AlEryani2019, Clerckx2018, Morsi2019, Varasteh2019, Varasteh2019b, Shanin2019}.
	
	\gls*{swipt} was studied first in \cite{Varshney2008}. 
	The author showed that there exists a fundamental trade-off between the achievable information rate and the transferred power in discrete-time memoryless Gaussian channels. 
	This trade-off is characterized by a non-increasing concave capacity-energy function. 
	{Further, in \cite{Zhang2013}, two fundamental SWIPT system architectures were proposed. Specifically, the authors considered SWIPT systems, where the \gls*{eh} and the \gls*{ir} are equipped with a single antenna and separate antennas, respectively. 
	Receivers with a single antenna serving co-located IR and EH have been studied for single-user \cite[Section III]{Clerckx2019}, \cite{Kang2017, Zhou2013} and relay-based \cite{Pilanawithana2019}, \cite{Gautam2018} SWIPT systems, whereas SWIPT systems with separate antennas for IR and EH have been considered in \cite[Section IV]{Clerckx2019}, \cite{Boshkovska2016, Boshkovska2015, Xiong2017, Zhou2013, Morsi2019, Varasteh2019, Varasteh2019b}.} 
	
	An essential prerequisite for the design of a \gls*{swipt} system is to accurately model the \gls*{eh} circuit which is employed to convert the received RF signal to a \gls*{dc} signal. 
	An \gls*{eh} circuit typically includes a rectenna, i.e., an antenna followed by a rectifier. In \cite{Varshney2008, Zhang2013, Pilanawithana2019, Gautam2018}, the authors assumed a linear relationship between the harvested power and the received \gls*{rf} power. 
	However, recently, practical non-linear models for \gls*{eh} circuits were proposed for the optimization of SWIPT systems \cite{Clerckx2018, Morsi2019, Boshkovska2016, Boshkovska2015, Varasteh2019, Varasteh2019b, Ma2019, Kang2017, Kang2018, Xiong2017, Kang2020, Zhou2013, AlEryani2019}.
	We note that for high input powers, typical EH circuits are driven into saturation by the diode breakdown effect, see, e.g., \cite[Figure 3]{Boshkovska2016}.
	{In \cite{Kang2018} and \cite{Ma2019}, power splitting between several rectifiers was proposed to mitigate saturation, which however leads to a more complicated EH design.} 
	The saturation behavior of the EH was modeled in \cite{Boshkovska2015} based on a sigmoid function, whose parameters are obtained by fitting the parameterized model to experimental data.
	{A similar model for the EH circuit was assumed in \cite{Kang2017}, where the authors determined the rate-energy trade-off of a SWIPT system employing a single-antenna receiver architecture.
	The analysis in \cite{Kang2017} was further extended to a SWIPT system with multiple antennas at the \gls*{tx}, \gls*{ir}, and \gls*{eh} in \cite{Xiong2017} and ergodic fading channels in \cite{Kang2020}.} 
	{Furthermore, in \cite{Clerckx2018, Zhou2013, AlEryani2019}, the authors investigated a non-linear diode model based on a Taylor series approximation of the current flow. 
	In \cite{Zhou2013} and \cite{AlEryani2019}, for \gls*{swipt} systems with single-antenna receivers and wireless powered communication systems, respectively, the authors neglected the higher-order terms in the Taylor series expansion such that the EH model reduced to the linear model in \cite{Varshney2008, Zhang2013, Pilanawithana2019, Gautam2018}, whereas, in \cite{Clerckx2018}, for multi-carrier transmission, the author showed that the input signal distributions maximizing the information rate and the transferred energy are different if the higher-order terms are not neglected.} 
	By varying the input distribution, different points in the corresponding rate-energy region can be achieved.
	In \cite{Morsi2019}, the authors considered a rectenna circuit comprising a single diode for signal rectification, developed a non-linear diode model for this circuit, and characterized the corresponding rate-energy region by optimally designing the input distribution to maximize the mutual information between the \gls*{tx} and the \gls*{ir} under a constraint on the minimum harvested power. 
	The analysis in \cite{Morsi2019} showed that the harvested power is maximized by allocating non-zero probabilities to two mass points which correspond to the maximum and minimum transmit power values and satisfy the \gls*{pp} and \gls*{ap} constraints of the \gls*{tx}.
	Finally, in \cite{Varasteh2019} and \cite{Varasteh2019b}, based on the concept of autoencoder \cite{Goodfellow2016}, the authors used a learning approach to optimize the modulation scheme to determine the trade-off between the symbol error rate for data transmission and the harvested power. 
	In \cite{Varasteh2019}, the design of the modulation scheme was based on the \gls*{eh} model in \cite{Morsi2019}, whereas in \cite{Varasteh2019b}, the authors utilized two different \gls*{eh} circuit models, one similar to the model in \cite{Clerckx2018} and one based on the model in \cite{Boshkovska2015}, for low and high \gls*{eh} input powers, respectively. 
	Similarly to \cite{Morsi2019}, the results in \cite{Varasteh2019b} suggest that on-off signaling is optimal for power transfer. 	
	
	Although the non-linear \gls*{eh} models considered in \cite{Clerckx2018, Morsi2019, Boshkovska2016, Boshkovska2015, Varasteh2019, Varasteh2019b, Ma2019, Kang2017, Kang2018, Xiong2017, Kang2020, Zhou2013, AlEryani2019} constitute a significant progress compared to the linear model in \cite{Varshney2008, Zhang2013, Pilanawithana2019, Gautam2018}, they are still based on strong assumptions. 
	First, it is assumed that the instantaneous harvested power depends on the currently received signal only. 
	However, rectifier circuits typically include a reactive element, usually a capacitor, as part of a \gls*{lp}. Since the voltage (or current) level on this element cannot drop instantaneously \cite{Horowitz1989}, rectenna circuits have \emph{memory}. 
	{Furthermore, for high RF signal powers, EHs suffer from the diode breakdown effect which was only partially analyzed in \cite{Morsi2019, Kang2018, Boshkovska2016, Boshkovska2015, Varasteh2019b, Kang2017, Kang2020, Xiong2017, Ma2019} and completely neglected in \cite{Clerckx2018, Varasteh2019, AlEryani2019, Zhou2013}.
	Moreover, the impedance values of the antenna and the rectifier have to be properly matched by a matching circuit (MC) in order to maximize the efficiency of the EH. This MC was assumed to be ideal in \cite{Clerckx2018, Boshkovska2016, Boshkovska2015, Morsi2019, Varasteh2019, Varasteh2019b, AlEryani2019, Zhou2013, Kang2017, Kang2018, Kang2020, Xiong2017, Ma2019}. 
	}	
	However, because of the rectifier non-linearity, perfect matching is possible for a single input signal frequency and a single power value only \cite{Mindan2010}, \cite{Tietze2012}. 
	Finally, in \cite{Clerckx2018, Morsi2019, Boshkovska2016, Boshkovska2015, Varasteh2019, Varasteh2019b} and other related works, the authors considered a rectenna circuit that comprised a single diode for half-wave signal rectification. 
	However, other rectifier circuits may be beneficial for \gls*{swipt} system design, e.g., the full-wave rectifier based on a bridge configuration with multiple diodes has been shown to lead to smaller output ripple \cite{Heljo2013} and higher diode breakdown voltage levels \cite{Horowitz1989}.
	
	In this paper, we develop an analytical framework for \gls*{swipt} system design and maximization of the rate-power region by optimization of the input signal distribution taking into account the above mentioned effects that were ignored in previous works. 	 
	{The main contributions of this paper can be summarized as follows:
	\begin{itemize}
		\item Since the behavior of an electrical circuit is determined by the initial state of its reactive elements and the input signal, we model the EH circuit by a discrete-time \gls*{mdp} \cite{Altman1999} which is a widely used framework to model and control systems with memory.
		{For SWIPT systems, MDPs were recently utilized to model the temporal correlation of wireless channels \cite{Kang2020a, Chun2018, Guo2020}. 
		Furthermore, in \cite{Pilanawithana2019} and \cite{Ku2015}, the authors considered relay-based systems and used MDPs to determine the transmit power at the relay nodes depending on time-varying system parameters.}
		In contrast to these works, we utilize an MDP to investigate the impact of the memory introduced by the reactive elements of practical EH circuits on the performance of SWIPT systems.
		\item Based on the MDP model, we study a SWIPT system, where \gls*{ir} and \gls*{eh} are equipped with separate antennas, and consider the cases where TX and IR have and do not have perfect knowledge of the instantaneous EH state. For both cases, we optimize the input signal distribution for maximization of the harvested power under constraints on the minimum mutual information between TX and IR and the maximum AP and PP at the TX. 	
		While, for the case, where TX and IR have perfect knowledge of the EH state, the formulated problem is convex, in the other case, the optimization problem is non-convex and we exploit alternating optimization \cite{Gorski2007, Grippo2000, Yu2019} to develop a low-complexity algorithm which is guaranteed to find a limit point of the problem.	
		We note that while, in practice, it may be difficult for TX and IR to track the EH state, the obtained boundary of the corresponding rate-power region can serve as a performance upper bound for SWIPT systems where this is not possible.	
		\item {Since an analytical model for the \gls*{eh} circuit that includes all non-linear effects of the rectifier and impedance mismatch is not tractable, we propose a learning based approach to deal with the non-idealities of the \gls*{eh} circuit. 
		In particular, we utilize dense neural networks (DNNs) to estimate the state transition probabilities and the immediate reward of the \gls*{mdp}.}
		\item {Our simulation results reveal that knowledge of the \gls*{eh} state at \gls*{tx} and \gls*{ir} can improve \gls*{swipt} system performance.	
		Moreover, our results show that the optimal input distribution and the rate-power region depend on the symbol duration, the \gls*{eh} impedance mismatch, the \gls*{eh} input signal power, and the type of rectifier circuit. 	
		In particular, a shorter symbol duration increases the achievable bit rate at the expense of a decrease of the average harvested power. 
		Furthermore, the half-wave rectifier is shown to yield a larger rate-power region in the low and medium input power regimes, whereas the full-wave rectifier is beneficial if the input power level at the \gls*{eh} is high.} 
	\end{itemize}
	}

	The rest of the paper is organized as follows. In Section \ref{SectionSystemModel}, we introduce the system model, propose the \gls*{mdp} model for the \gls*{eh}, and discuss the information transmission to the \gls*{ir}. 
	In Section \ref{SectionProblemFormulation}, to determine the boundary of the rate-power region of the considered \gls*{swipt} system, we formulate two optimization problems for the cases where both \gls*{tx} and \gls*{ir} know and do not know the instantaneous \gls*{eh} state, respectively.
	In Section \ref{SectionNumericalResults}, we provide simulation results for performance evaluation.
	Finally, in Section \ref{SectionConclusion}, we draw some conclusions.
	
	Throughout this paper, we use the following notations. Bold lower case letters stand for vectors, i.e., $\boldsymbol{x}$ is a vector, and its $i^\text{th}$ element is denoted by ${x}_i$. Bold upper case letters represent matrices, i.e., $\boldsymbol{X}$ is a matrix and ${X}_{i,j}$ is its element in the $i^\text{th}$ row and $j^\text{th}$ column. 
	The average value of \gls*{rv} $x$ is denoted by $\overline{x}$. $f(x,y;z)$ denotes a function of variables $x$ and $y$ for a given parameter $z$.
	The indicator function $\boldsymbol{1}_{X}(x)$ takes value $1$ if $x$ is in set $X$ and $0$ otherwise. 
	{ The exponential function of $x$ is denoted by $e^x$.} 
	$\mathbb{E}_x\{\cdot\}$ denotes the expectation with respect to the distribution of \gls*{rv} $x$. 
	Operator $\Re\{\cdot\}$ denotes the real part of a complex number. 
	The Moore-Penrose pseudoinverse of a matrix is denoted by $(\cdot)^\dagger$. 
	$\mathbb{R}$ and $\mathbb{C}$ indicate the sets of real and complex numbers, respectively. 
	The imaginary unit is denoted by $j$. 
	The circularly-symmetric complex Gaussian distribution with mean vector $\boldsymbol{\mu}$ and covariance matrix $\boldsymbol{\Gamma}$ is denoted by $\mathcal{CN}(\boldsymbol{\mu}, \boldsymbol{\Gamma})$. 
	$\mathrm{Pr} \{x = {x}_{i}\}$ stands for the probability that \gls*{rv} $x$ is equal to a particular value ${x}_{i}$.

	\section{System Model and Preliminaries}
	\label{SectionSystemModel}
	In this section, first, we present the considered \gls*{swipt} system model. Then, we model the \gls*{eh} by an \gls*{mdp} and discuss the information transmission to the \gls*{ir}.	
	\subsection{System Model}
\begin{figure}[!t]
		\centering
		\includegraphics[width=0.45\textwidth]{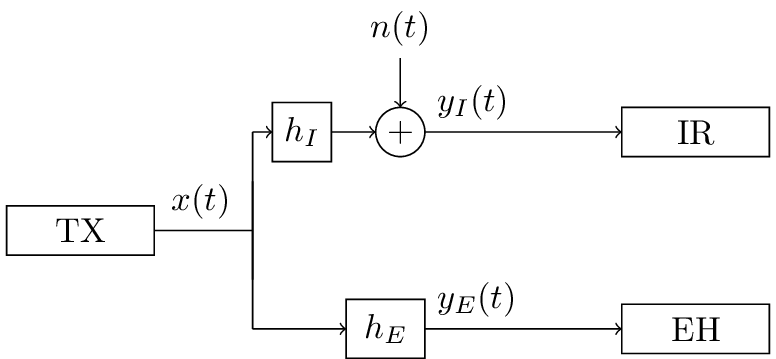}
		
		\caption{SWIPT system model comprising a \gls*{tx}, an \gls*{ir}, and an \gls*{eh}.}
		\label{System_Fig}
	\end{figure}

	Let us consider a SWIPT system where the \gls*{tx}, \gls*{ir}, and \gls*{eh} each have a single antenna\footnotemark, see Fig. \ref{System_Fig}. The \gls*{tx} broadcasts a pulse-modulated signal, which is modeled as $x(t) = \sum_{k=0}^{\infty}x[k] \psi (t-kT)$, where $T$ is the symbol duration, $\psi(t)$ is a rectangular transmit pulse shape, and $x[k] \in \mathbb{C}$ is the information symbol transmitted in time slot $k$.   	
	\footnotetext{As in \cite[Section IV]{Clerckx2019}, \cite{Morsi2019}, and references therein, in this paper, we assume that the EH and IR are separate devices.
	Our framework can be applied to SWIPT systems with spatially separated or co-located EH and IR as long as both devices are equipped with their own antenna, respectively.}
	
	The complex-valued information symbol $x[k], k \in \{0, 1, ...\}$, is expressed in polar coordinates as $x[k] = r_x[k] e^{j \phi_x[k]}$, where the amplitude $r_x[k] \geq 0$ is a realization of an \gls*{iid} \gls*{rv} $r_x$, whereas the phase $\phi_x[k] \in [-\pi, \pi)$ is a realization of an \gls*{iid} \gls*{rv} $\phi_x$. We denote the joint \gls*{pdf} of \gls*{rv}s $r_x$ and $\phi_x$ by $p_{r_x, \phi_x} (r, \phi)$. 
	{The complex-valued fading gains of the \gls*{ir} and \gls*{eh} channels are assumed to be constant within a coherence time interval and are given by $h_I = |h_I| e^{j \phi_I}$ and $h_{E} = |h_{E}| e^{j \phi_{E}}$, respectively.}
	We assume that the \gls*{tx} has perfect knowledge of both channel gains and, additionally, $h_I$ is known at the \gls*{ir}. 
	The RF signals received at the \gls*{ir} and the \gls*{eh} can be expressed as $y_{I}^{RF}(t) = \sqrt{2}\Re\{[h_I x(t) + n(t)] e^{j 2 \pi f_ct}\}$ and $y_{E}^{RF}(t) = \sqrt{2}\Re\{h_{E} x(t) \, e^{j 2 \pi f_ct}\}$, respectively, where $f_c$ and $n(t)$ denote the carrier frequency and complex-valued zero-mean \gls*{awgn}, respectively. 
	We note that the noise received at the \gls*{eh} is ignored because its contribution to the harvested energy is negligible.
	\subsection{MDP Model of \gls*{eh}}
	\label{SectionEnergyHarvester}
	
	In this section, we develop an \gls*{mdp} model for the \gls*{eh}.
	An \gls*{mdp} consists of a finite state space of size $S_\Xi$, $\Xi~=~\{\xi_1,~\xi_2,~...,~\xi_{S_\Xi}\}$, a set of actions, $\mathcal{X}_E$, transition probabilities, and a reward function. 
	In particular, the current \gls*{mdp} state $\xi[k] \in \Xi$ depends on the previous state $\xi[k-1] \in \Xi$ and action $x_E[k] \in \mathcal{X}_E$ only. Moreover, the transition from state $\xi_i \in \Xi$ to state $\xi_j \in \Xi$ due to action $x_E$ occurs with transition probability $\hat{\rho}_{i,j}(x_E)$, and, during this transition, an immediate reward $\hat{P}_{i,j}(x_E)$ is received. 	
	
	In the following, first, we present the rectenna circuit employed by the \gls*{eh}. Then, we provide the \gls*{mdp} model for the \gls*{eh}. Finally, we derive an expression for the average power harvested by the \gls*{eh} which represents the average reward of the \gls*{mdp}. 
	
	\subsubsection{\gls*{eh} Circuit}
\begin{figure}[t]
		\centering		
		\includegraphics[width=0.48\textwidth]{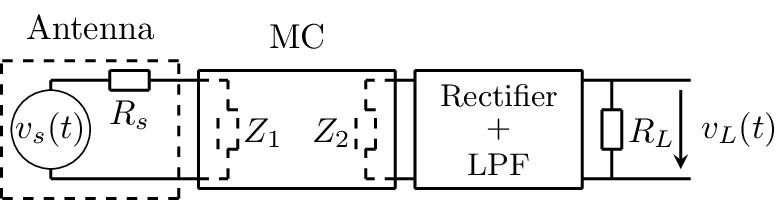}
		
		\caption{\gls*{eh} circuit model comprising an antenna, an MC, a rectifier with an \gls*{lp}, and a load resistor $R_L$.}
		\label{EH_Fig}	
	\end{figure}

	Similar to \cite{Clerckx2019, Morsi2019}, and references therein, we assume that the \gls*{eh} is equipped with a non-linear rectenna circuit comprising an antenna, an MC, a rectifier with an \gls*{lp}, and a load resistor, cf. Fig. \ref{EH_Fig}. 
	The antenna is modeled as a voltage source $v_s(t)$ connected in series with resistance $R_s$.	
	The rectifier circuit typically includes a non-linear diode circuit and an \gls*{lp} to convert the \gls*{rf} signal received at the \gls*{eh} to a low frequency output voltage $v_L(t)$ across the load resistance $R_L$. 		
	Finally, as in \cite{Morsi2019} and \cite{LePolozec2016}, the \gls*{eh} includes an impedance MC, which matches the antenna output impedance $Z_1$ and the input impedance of the rectifier circuit $Z_2$ to maximize the power transferred from the antenna to the rectifier. 
	Note that since the rectifier circuit includes non-linear elements, typically diodes, exact matching is possible for one frequency and one power value of the received signal only.	
	Examples for the employed rectifiers and MCs will be provided in Section \ref{SectionNeuralNetwork}.
	
	Since the output voltage of the rectenna circuit $v_L(t)$ is always bounded above due to the diode breakdown effect \cite{Guo2014}, we define the maximum load voltage level as $V_L^\text{max}$. 
	We introduce a finite set of \emph{voltage levels} of size $S_\Xi + 1$, whose elements are defined as $\hat{v}_{l} = \frac{V_L^\text{max}}{S_\Xi}l$, $l \in \{0,1,...,S_\Xi\}$. 
	Furthermore, we define the set of \emph{quantized load voltage} levels of size $S_\Xi$, whose elements are given by $\tilde{v}_i = \frac{\hat{v}_{i-1} + \hat{v}_{i}}{2}$, $i \in \{1,2,...,S_\Xi\}$.
	We approximate the output voltage level $v_L(t)$ by discrete value $\tilde{v}_i$ if $v_L(t)~\in~[\hat{v}_{i-1}, \hat{v}_i)$. 
	Note that if the number of quantization levels approaches infinity, i.e., $S_\Xi \to \infty$, then $\tilde{v}(t) \to v_L(t)$.
	
	\subsubsection{\gls*{eh} States} 
	\label{SectionEhStates}
	In the following, we model the \gls*{eh} as an \gls*{mdp}.	
	For the proposed model, we treat the quantized voltage levels at the load resistor, $\tilde{v}_l, l \in \{1,2,...,S_\Xi\}$, and the received symbols at the \gls*{eh}, $x_E[k] = h_E x[k]$, as the states and actions of the \gls*{mdp}, respectively.
	Furthermore, we approximate the amount of power harvested during the transition from the previous \gls*{eh} state $\xi$ to the current state due to received symbol $x_E$ by the immediate reward function $P(\xi, x_E)$.		
	
	More in detail, we map the discrete voltage levels $\tilde{v}$ to the \gls*{eh} states $\xi \in \Xi$ of a stochastic process, where $\Xi \subset \mathbb{R}^{S_\Xi}$ is a finite state space\footnotemark.
	\footnotetext{In this work, we consider EH circuits comprising a first-order LPF, which includes a single reactive element. 
	We note that the proposed framework can be extended to general EH circuits with multiple reactive elements as part of, e.g., a higher-order LPF \cite{Tietze2012} or the model of a battery \cite{Wehbe2015}, where the harvested energy may be stored for future use \cite{Sekander2020}, by increasing the dimensions of the MDP state space \cite{Feng2012}. 
	In this case, the general MDP state is an $m$-tuple $\{\xi^1,..., \xi^m\} \in \Xi^1 \times ... \times \Xi^m$, where $m$ is the number of reactive elements and $\xi^i \in \Xi^i, i\in\{1,2, ..., m\}$ represents the voltage (or current) level at the $i^\text{th}$ reactive element.}
	We note that since the received RF signal $y_{E}^{RF}(t)$ is time-slotted, so is the output voltage signal $v_L(t)$.
	Thus, the system is in state $\xi[k]=\xi_i \in \Xi$, if at time instant $t=kT$, the load voltage level $v_L(kT) \in [\hat{v}_{i-1}, \hat{v}_i)$, such that it is approximated by the quantized voltage level $\tilde{v}(kT)$ associated with this state, i.e., $\xi[k] = \xi_i = \tilde{v}_i= \tilde{v}\big(kT\big)$. 	
	{Furthermore, we assume that when the EH is in state $\xi_i$, each voltage level from the range $[\hat{v}_{i-1}, \hat{v}_i)$ is equiprobable. This assumption is justified if the number of quantization levels is large.}
	Moreover, due to the memory introduced by the \gls*{lp}, at the end of time interval $k$, the attained value $v_L(kT)$ of the load voltage level depends on the received symbol $x_{E}[k] = h_{E} x[k]$ and the load voltage level at the end of the previous time interval, $v_{L}((k-1)T)$.	
	We note that although the MC typically includes additional reactive elements, it is designed as a band-pass filter for the received signal and fine-tuned to carrier frequency $f_c$. 
	Furthermore, the bandwidth of this filter is much larger than the symbol rate $\frac{1}{T}$. 
	Therefore, the memory introduced by the MC to the energy harvesting process is negligible, and thus, we do not include the reactive elements of the MC in our \gls*{mdp} model.	
	
	Stochastic process $\{\xi[k]\}$ may change its value in each symbol interval, i.e., when the \gls*{eh} receives a new symbol. 
	Therefore, $\{\xi[k]\}$ is a discrete-time process and its \emph{time step} is equal to the symbol duration.	
	Thus, since the behavior of the \gls*{eh} circuit in a given time interval is completely determined by the initial conditions and the input signal \cite{Horowitz1989}, $\xi[k]$ depends only on the voltage level of the load resistance at time $(k-1)T$, i.e., the previous state $\xi[k-1]$, and the received symbol $x_E[k]$. 
	Hence, the symbol $x_E[k]$ received at the \gls*{eh} corresponds to an action of the \gls*{mdp} that is taken in time step $k$, i.e., when the \gls*{mdp} transits from state $\xi[k-1]$ to state $\xi[k]$.
	Thus, the probability of any state of the stochastic process $\xi[k]$ depends only on the previous state $\xi[k-1]$ and the received symbol\footnote{In this work, as in \cite{Morsi2019, Clerckx2018}, and references therein, we assume a rectangular pulse shape for the transmit filter. Therefore, there is no interference between consecutive received symbols at the EH antenna. 
	We note that if, for example, a Root-Raised Cosine pulse shaping filter is employed at the TX, the intersymbol interference caused at the EH introduces additional memory. In particular, in this case, $\xi[k]$ depends also on $x_E[k+i], i \neq 0$, which influences the power harvested at the EH.	
	The study of the impact of the memory introduced by pulse shaping filters that span several symbol intervals on system performance is beyond the scope of this work. Nevertheless, we note that \gls*{swipt} systems employing such pulse shaping filters can be studied by increasing the dimension of the Markov process \cite{Chen_2008, Feng2012}.} $x_{E}[k]$, i.e., $\mathrm{Pr} \{ \xi[k]\mid x_{E}[k],\xi[{k~-~1}], x_{E}[k~-~1], \xi[{k-2}], \, ..., \, x_{E}[1], \xi[{0}] \} = \mathrm{Pr} \{ \xi[k] \mid x_{E}[k], \xi[k-1]\}$.	 
	Furthermore, we note that for the considered narrowband signals, the symbol duration is typically much larger than the period of the \gls*{rf} signal, i.e., $T \gg \frac{1}{f_c}$.
	Moreover, the time constant of the \gls*{lp} is typically also much larger than $\frac{1}{f_c}$.
	Hence, we assume that the rectifier behaves as an envelope detector \cite{Tietze2012}, and hence, neglect the influence of the phase changes from one received symbol to the next one on the harvested power. 
	Therefore, the state $\xi$ is independent of the phase $\phi_{x_{E}} = \phi_x+\phi_{E}$ of the received signal and depends on its amplitude $r_{E} = |h_{E}| r_x$ only. 
	Moreover, since channel gain $h_{E}$ is assumed to be perfectly known at the \gls*{tx}, $\mathrm{Pr} \{ \xi[k] \mid r_x[k], \xi[k-1]\} = \mathrm{Pr} \{ \xi[k] \mid x_{E}[k], \xi[k-1]\}$ and, thus, the sequence of pairs $\{\xi, r_x\}$ can be modeled as an \gls*{mdp} \cite{Norris1998, Meyn2012}.
	
	We denote the probability of transition from state $\xi_i \in \Xi$ to state $\xi_j \in \Xi$ when a symbol with amplitude $r_x$ is transmitted by $\rho_{i,j}(r_x) = \mathrm{Pr} \{ \xi_j \mid r_x, \xi_i\}$. 
	Next, we note that for a given $v_L\big((k-1)T\big) = v_\nu$, the reception of a symbol with amplitude $r_E = |h_E| r_x$ determines the output voltage level in the next time slot, $v_L\big(kT\big) = v_{\mu}$. 
	Then, $\mathrm{Pr} \{ v_{\mu} \mid r_E, v_{\nu} \}$ is equal to 1 if the reception of a symbol with amplitude $r_E$ leads to the transition from $v_{\nu}$ to $v_{\mu}$ and 0, otherwise.
	Thus, $v_\mu$ can be obtained as a deterministic function of $r_E$ and $v_\nu$, i.e., $v_\mu = f_v(v_{\nu}, r_E)$.
	However, since the states $\xi_i$ and $\xi_j$ comprise voltage levels from the intervals $\hat{V}_i = [\hat{v}_{i-1}, \hat{v}_i)$ and $ \hat{V}_j = [\hat{v}_{j-1}, \hat{v}_j)$, respectively, the transition probabilities of the discrete \gls*{mdp} $\rho_{i,j}(r_x)$ may take any value from the interval $[0,1]$ and can be calculated as follows
	\begin{equation}
	\rho_{i,j}(r_x) = \frac{ \int_{\hat{V}_i } \mathbf{1}_{\hat{V}_j} \big( f_v(v, |h_E| r_x) \big) dv }{ \hat{v}_i - \hat{v}_{i-1} }.
	\label{TransitionProbabilityEqn}
	\end{equation}
	\noindent We note that an analytical evaluation of $\rho_{i,j}(r_x)$ is intractable since the transition function $f_v(v_{\nu}, r_E)$  cannot be derived analytically due to the rectifier non-linearity, the imperfections of the MC, and the circuit memory. 
	Therefore, in Section \ref{SectionNeuralNetwork}, we employ a \gls*{dnn} \cite{Goodfellow2016} to approximate $f_v(v_\nu, r_{E})$ and, thus, to compute the transition probabilities $\rho_{i,j}(r_x)$.

	{Finally, we assume that the \gls*{mdp} is unichain and ergodic, i.e., for a given input signal distribution, starting from any initial state, the \gls*{mdp} always reaches the same steady-state distribution \cite{Norris1998}.}
	Then, we denote the joint \gls*{pdf} of state $\xi_i \in \Xi$ and symbol amplitude $r_{x}$ by $\boldsymbol{\pi}_i(r_x)$ and collect these \gls*{pdf}s in vector $\boldsymbol{\pi}(r_x) \in \mathbb{R}^{S_\Xi}$, which is a solution of the following system of balance equations \cite{Altman1999}
	\begin{equation}
	\label{FiniteBalanceEqn1}
	\begin{aligned}
		\sum_{i=1}^{S_\Xi} \int_{r_x} {\pi}_{i}(r_x) \big( \mathbf{1}_j(i) - \rho_{i,j}x(r_x) \big) & dr_x = 0, \\[-1.2ex]
		&\forall j \in \{1,2,...S_\Xi\},
	\end{aligned}
	\end{equation}	
	\begin{equation}
	\label{FiniteBalanceEqn2}
	\sum_{i=1}^{S_\Xi} \int_{r_x} {\pi}_{i}(r_x)dr_x = 1.
	\end{equation}	
		
	\subsubsection{\gls*{mdp} Reward}
	\label{SectionAveragePower}
	In the following, we derive an expression for the average harvested power which constitutes the average reward of the \gls*{mdp}.
	
	First, we note that the instantaneous harvested power can be expressed as ${P(t)} =  \frac{v_L^2(t)}{R_L}$ and, thus, similarly to $v_L(t)$, $P(t)$ is a time-slotted random process. 
	We denote the average power harvested while receiving an infinitely long sequence of random symbols $\{x_{E}[k]\}$ by $\overline{P}$. This value can be estimated by averaging function $P(t)$ over time, or equivalently, over time intervals, assuming that the number of time intervals $K$ approaches infinity:
	\setlength{\arraycolsep}{0.0em}
	\begin{equation}
		\begin{aligned}
			\overline{P} &= \lim_{t \rightarrow \infty} \frac{1}{t} \int_{0}^{t} P(\tau) d\tau \\[-0ex]
			&= \lim_{K \rightarrow \infty} \frac{1}{K} \sum_{k=0}^{K-1} \frac{1}{T} \int_{0}^{T} P(t+kT) dt.
		\end{aligned}
	\label{EHCircuitEqn}
	\end{equation}	
	Since the amount of power harvested by the \gls*{eh} during time slot $k$ depends on the amplitude $r_{E} = |h_E| r_x$ of the symbol received in time slot $k$ and on the load voltage level at the end of the previous time slot $v_\mu = v_L\big((k-1)T\big)$, we define the harvested power averaged over symbol duration $T$ by $P'(v_\mu, r_{E}) = \frac{1}{T} \int_{0}^{T} P\big(t+(k-1)T\big) dt$. 
	Furthermore, the average amount of power, $\tilde{P}_i(r_{x})$, harvested if the previous state is $\xi[k-1] = \xi_i \in \Xi$ and a symbol with amplitude $r_E = |h_E| r_x$ is received, is approximated as $\tilde{P}_i(r_{x}) = P'(\tilde{v}_i, |h_E| r_x)$.
	Additionally, since the \gls*{mdp} is assumed to be ergodic, the influence of the initial state on the average harvested power vanishes as the length of the symbol sequence $\{x[k]\}, k = \{0,1,2,...\}$, increases. 
	Hence, the average harvested power can be obtained as a function of $\boldsymbol{\pi}(r_x)$ by averaging $\tilde{P}_i(r_{x})$ over both the \gls*{eh} states and the transmitted symbols as follows\footnotemark
	\begin{align}
		\overline{P}( \boldsymbol{\pi} ) = \mathbb{E}_{\xi, r_{x}} \big\{ \tilde{P}_i(r_{x}) \big\} = \sum_{i=1}^{S_\Xi} \int_{r_x} {\pi}_i(r_x) \tilde{P}_i(r_{x}) d r_x .
		\label{ExpandedAverageRewardEqn}
	\end{align}		
	\footnotetext{From our simulations (not included in the paper), we observed that because of the unichain ergodicity of the MDP, for a valid pdf $\boldsymbol{\pi}(r_x)$, the time averaged harvested power in (4) converges to the value $P(\boldsymbol{\pi})$ in (5) after a finite transient time (e.g., $K = 5000$ time slots), which we assume to be sufficiently short compared to the coherence time interval of the EH channel.} 
	Since ${P}'(v, r_{E})$ is also not analytically tractable, in Section \ref{SectionNeuralNetwork}, we employ a second \gls*{dnn} to approximate this function.	

	\subsection{Information Receiver}
	\label{SectionInformationReceiver}	
	Let us consider the signal received at the \gls*{ir}, $y_{I}^{RF}(t)$. 
	Since $y_{I}^{RF}(t)$ is a time-slotted signal, after down-conversion, matched filtering, and sampling, the received signal in time slot $k$ can be expressed as $y[k] = h_I x[k] + n[k]$, where $n[k]$ is discrete-time \gls*{awgn} distributed as $\mathcal{CN}(0,2\sigma^2_n)$. 
	Furthermore, $y[k]$ can be expressed in polar coordinates as $y[k] = r_y[k] e^{j \phi_y[k]}$ with $r_y[k] \geq 0$ and $\phi_y[k] \in [-\pi, \pi)$, where amplitude $r_y[k]$ and phase $\phi_y[k]$ are realizations of \gls*{iid} \gls*{rv}s $r_y$ and $\phi_y$, respectively. 
	We denote the joint \gls*{pdf} of \gls*{rv}s $r_y$ and $\phi_y$ as a function of the joint input \gls*{pdf} $p_{r_x, \phi_x}$ by $p_{r_y, \phi_y}\big(r,\phi; p_{r_x, \phi_x}\big)$.
	
	The mutual information between $x$ and $y$ as a function of the joint input \gls*{pdf} $p_{r_x, \phi_x}(r, \phi)$ can be expressed as \cite{Smith1971}	
	\begin{equation}
	I\big(p_{r_x, \phi_x} \big) = H_y \big(p_{r_x, \phi_x} \big) - H_n,
	\end{equation}
	\noindent where $H_y \big(p_{r_x, \phi_x} \big)$ and $H_n$ are the differential entropies of the received signal and the noise, respectively. 
	We note that the differential entropy of the noise does not depend on the input \gls*{pdf} $p_{r_x, \phi_x}(r, \phi)$ and is equal to $H_n = \text{log}_2(2\pi e \sigma_n^2)$ \cite{Tse2005}. 
	The differential entropy of the complex-valued received signal can be expressed as follows \cite{Shamai1995}
	\begin{equation}
		\begin{aligned}	
			H_y\big( p_{r_x, \phi_x} \big) &= H_{r_y,\phi_y}\big( p_{r_x, \phi_x} \big) \\
			&\quad + \int_{0}^{\infty} p_{r_y}\big(r; p_{r_x, \phi_x}\big) \, \text{log}_2 (r) \, dr,
		\end{aligned}
	\end{equation}
	\noindent where $p_{r_y}\big(r; p_{r_x, \phi_x}\big)$ and $H_{r_y,\phi_y}\big( p_{r_x, \phi_x} \big)$ denote the \gls*{pdf} of \gls*{rv} $r_y$ and the differential entropy of \gls*{rv}s $r_y$ and $\phi_y$ as functions of input \gls*{pdf} $p_{r_x, \phi_x}$, respectively.

	\section{Rate-Power Region of SWIPT System}
	\label{SectionProblemFormulation}		
	We refer to the set of all attainable pairs of average harvested powers and achievable rates as the rate-power region of the SWIPT system \cite{Morsi2019}. 
	In order to obtain the boundary of this rate-power region, in this section, we jointly optimize the \gls*{pdf}s of the \gls*{eh} states $\xi$, transmit symbol amplitudes $r_x$, and transmit symbol phases $\phi_{x}$ for maximization of the harvested power at the \gls*{eh} under a constraint on the minimum required mutual information between \gls*{tx} and \gls*{ir}.
	To this end, we first consider the case where both \gls*{tx} and \gls*{ir} have perfect knowledge of the instantaneous \gls*{eh} state which leads to a convex optimization problem.  
	Then, we consider the more practical case, where the \gls*{eh} state is not known at \gls*{tx} and \gls*{ir}. In this case, the resulting optimization problem is non-convex, and we develop an iterative algorithm to obtain a limit point \cite{Yu2019}.
	
	\subsection{\gls*{eh} State Is Known at \gls*{tx} and \gls*{ir}}
	\label{SectionColocatedReceivers}
	\setcounter{equation}{9}
	\begin{figure*}[!t]
		\normalsize
		\begin{equation}
		I\big( p^i_{r_x, \phi_x} \big) = I\big( p^i_{r_x} \big)= -\int_{0}^{\infty} p_{r_y}\big(r_y; p^i_{r_x}\big) \log_2 \Big( \frac{1}{r_y} p_{r_y}\big(r_y; p^i_{r_x}\big) \Big)  dr_y + \log_2 (2 \pi) - H_n,
		\label{ProposedMutualInf}
		\end{equation}
		\begin{equation}
		p_{r_y}\big(r_y; p^i_{r_x}\big) = \frac{1}{\sigma_n^2} \int_{r_x} r_y e^{-\frac{r_y^2 + r_x^2 |h_I|^2}{2 \sigma_n^2}} I_0 \bigg( \frac{r_y \, r_x \, |h_I|}{\sigma_n^2} \bigg) p^i_{r_x}(r_x) dr_x.
		\label{ProposedOutputPdf}
		\end{equation}
		\vspace*{4pt}
		\hrulefill
		\vspace*{8pt}
	\end{figure*}	
	\setcounter{equation}{7}
	{In this section, we formulate a convex optimization problem for the joint pdf of the EH states $\xi$, transmit symbol amplitudes $r_x$, and phases $\phi_x$ for the case when the EH state is known at the TX and the IR. Then, we show that for the solution of the problem, the phase $\phi_x$ is statistically independent from $\xi$ and $r_x$ and uniformly distributed. Finally, exploiting this observation, we reformulate and solve the optimization problem.} 
	
	In this section, we assume that both \gls*{tx} and \gls*{ir} know the current \gls*{eh} state. 
	This assumption may hold in practice if, e.g., the \gls*{ir} and \gls*{eh} are co-located devices \cite{Clerckx2019} and the \gls*{tx} is able to track the \gls*{eh} state.	
	Under this assumption, the \gls*{pdf} of the transmit symbols, which must be known at \gls*{tx} and \gls*{ir}, can be made dependent on the \gls*{eh} state $\xi$ and, hence, be modeled as $p^i_{r_x, \phi_x} = p^i_{r_x} p^i_{\phi_x|r_x}$, where subscript $i$ refers to the \gls*{pdf} for \gls*{eh} state $\xi_i$ and $p^i_{\phi_x|r_x}$ is the conditional \gls*{pdf} of phase $\phi_x$ for a given amplitude $r_x$.	
	Thus, we obtain the boundary of the rate-power region by jointly optimizing the joint \gls*{pdf} of the \gls*{eh} states $\xi$ and the amplitudes $r_x$ of transmitted symbol $x$, $\boldsymbol{\pi}(r_x)$, and the set of conditional \gls*{pdf}s $\mathcal{P}_{\phi_x \mid r_x}$, whose $i^\text{th}$ element is $\mathcal{P}^i_{\phi_x|r_x} = p^i_{\phi_x|r_x}$, $i~\in~\{1,2,...,S_\Xi \}$. 
	
	We note that for a given joint \gls*{pdf} $\boldsymbol{\pi}(r_x)$, the \gls*{pdf} of the symbol amplitudes in state $\xi_i$ is given by \cite{Altman1999}	
		\begin{equation}	
		p_{r_x}^i(r) = \frac{{\pi}_{i}(r)}{{\gamma}_{i}},
		\label{PolicyEqn}
		\end{equation}		
	\noindent where ${\gamma}_{i} = \int_{r_x} {\pi}_{i}(r_x) d r_x$ denotes the marginal probability of state $\xi_i \in \Xi$. 	
	Additionally, since in each symbol interval $k$, transmitted symbol $x[k]$ may be taken from a different distribution depending on the current \gls*{eh} state, we introduce the \emph{expected mutual information} averaged over the \gls*{eh} states, which is given by $\overline{I}\big( \boldsymbol{\pi}, \mathcal{P}_{\phi_x|r_x} \big) = \sum_{i=1}^{S_\Xi} {\gamma}_i I\big( p^i_{r_x, \phi_x} \big)$.
	Hence, we formulate the following constrained optimization problem	
	\begin{subequations}
		\label{MajorOptProblem}
		\begin{align}
		\label{MajorOptProblem_Objective}
		\maximize_{\boldsymbol{\pi}(r_x), \mathcal{P}_{\phi_x \mid r_x}}  \; & \overline{P}( \boldsymbol{\pi} ) \\ 
		\text{subject to}\quad &
		\label{MajorOptProblem_Constr1}		
		\overline{I}\big(  \boldsymbol{\pi}, \mathcal{P}_{\phi_x|r_x} \big) 
		\geq I_{\text{req}}, \\
		\label{MajorOptProblem_Constr2}		
		& \sum_{i=1}^{S_\Xi} \int_{r_x} r_x^2 {\pi}_i(r_x) dr_x \leq \sigma_{r_x}^2, \\
		\label{MajorOptProblem_Constr3}
		& |r_x| \leq r_x^\text{max}, 
		\\
		\label{MajorOptProblem_Constr4}
		& \sum_{i=1}^{S_\Xi} \int_{r_x} {\pi}_{i}(r_x) d r_x = 1,
		\\	
		\label{MajorOptProblem_Constr5}
		&\begin{aligned} \sum_{i=1}^{S_\Xi} \int_{r_x} {\pi}_{i}(r_x) \big( \mathbf{1}_j&(i) - \rho_{i,j}(r_x) \big) d r_x \\[-1.2ex]%
		{}&= 0, \; j \in \{1,2,...S_\Xi\}, \end{aligned}
		\end{align}
	\end{subequations}	
	\noindent where $I_{\text{req}}$ in (\ref{MajorOptProblem_Constr1}) is the minimum required expected mutual information between \gls*{tx} and \gls*{ir}. 
	Constraint (\ref{MajorOptProblem_Constr2}) limits the \gls*{ap} budget at the \gls*{tx} to $\sigma^2_{r_x}$ to avoid excessive power consumption and interference to other systems.	
	Moreover, to avoid driving the power amplifier into a non-linear regime, we impose constraint (\ref{MajorOptProblem_Constr3}) to limit the \gls*{pp} at the \gls*{tx} by introducing the maximum amplitude $r_x^\text{max}$. 
	Furthermore, constraints (\ref{MajorOptProblem_Constr4}) and (\ref{MajorOptProblem_Constr5}) ensure that the solution $\boldsymbol{\pi}(r_x)$ is a valid \gls*{pdf}, i.e., summing the probabilities over the \gls*{mdp} state and action spaces yields one, and corresponds to the steady-state distribution of the \gls*{mdp} described in Section \ref{SectionEnergyHarvester}, respectively. 
	
	Note that since $\boldsymbol{\pi}(r_x)$ is independent of $\phi_{x}$, in (\ref{MajorOptProblem}), only $\overline{I}\big( \boldsymbol{\pi}, \mathcal{P}_{\phi_x|r_x} \big)$ depends on the distribution of the phases $\phi_x$. 
	In the following proposition, we show that for the solution of (\ref{MajorOptProblem}), the individual phase distributions in $\mathcal{P}_{\phi_x \mid r_x}$ are independent of \gls*{eh} state $\xi$, i.e., $p^1_{\phi_x|r_x} = p^2_{\phi_x|r_x} = ... = p^{S_\Xi}_{\phi_x|r_x}$, and the phases $\phi_x$ are statistically independent of the symbol amplitude $r_x$ and uniformly distributed.
		
	\begin{proposition}
		\label{Theorem1}
		For the solution of (\ref{MajorOptProblem}), the phase $\phi_x$ of transmitted symbol $x = r_x \, e^{j \phi_x}$ is uniformly distributed and statistically independent from amplitude $r_x$ and \gls*{eh} state $\xi$. Moreover, \gls*{rv}s $r_y$ and $\phi_y$ are also statistically independent and phase $\phi_y$ is uniformly distributed. Furthermore, the mutual information and the distribution of the amplitudes of the received symbol, when the \gls*{eh} is in state $\xi_i$, can be simplified as in (\ref{ProposedMutualInf}) and (\ref{ProposedOutputPdf}), shown on top of this page, respectively.
		\setcounter{equation}{11}				
		\begin{proof}
			Please refer to Appendix \ref{AppendixA}. 
		\end{proof}	
	\end{proposition}
	Exploiting Proposition \ref{Theorem1}, we can reformulate the constrained optimization problem in (\ref{MajorOptProblem}) as follows:
	
	\begin{subequations}
		\label{RefMajorOptProblem}
		\begin{align}
		\maximize_{\boldsymbol{\pi}(r_x)} \; & \overline{P}( \boldsymbol{\pi} ) \\
		\text{subject to}\quad &	
		\overline{I}( \boldsymbol{\pi}) = \sum_{i=1}^{S_\Xi} {\gamma}_i I\big( p^i_{r_x} \big) \geq I_{\text{req}},\\		
		& \text{(\ref{MajorOptProblem_Constr2})-(\ref{MajorOptProblem_Constr5})}, \nonumber
		\end{align}
	\end{subequations}	
	\noindent where the pdf of the symbol amplitudes, $p^i_{r_x}(r)$, is calculated as in (\ref{PolicyEqn}). 
	
	The solution of (\ref{RefMajorOptProblem}), $\boldsymbol{\pi}^*(r_x)$, is the optimal joint \gls*{pdf} of the \gls*{eh} states and the transmitted symbols and maximizes the average harvested power at the \gls*{eh} subject to the constraints on the expected mutual information between \gls*{tx} and \gls*{ir}, the \gls*{ap}, and the \gls*{pp} at the \gls*{tx}. 
	{Since the EH state and the EH circuit parameters are assumed to be known at the \gls*{tx}, convex optimization problem (\ref{RefMajorOptProblem}) can be efficiently solved at the \gls*{tx} node using standard numerical tools, such as CVX \cite{Grant2015}.}
	
	To implement the policy obtained by solving (\ref{RefMajorOptProblem}), the input distribution has to be adapted at the \gls*{tx} according to the instantaneous \gls*{eh} state, i.e., \gls*{tx} and \gls*{ir} must have perfect knowledge of the current \gls*{eh} state.
	This may be difficult to realize in practice.
	Nevertheless, $\boldsymbol{\pi}^*(r_x)$ constitutes a performance upper-bound for the more practical case where the \gls*{eh} state is not known at \gls*{tx} and \gls*{ir}. 
	The rate-power region for this case will be tackled next.
	
	\subsection{\gls*{eh} State Is Not Known at \gls*{tx} and \gls*{ir}}
	\label{SectionparatedReceivers}	
	Now, we consider \gls*{swipt} systems where the current \gls*{eh} state is not known at \gls*{tx} and \gls*{ir}. 
	Tracking the \gls*{eh} state increases complexity and may not be possible when \gls*{eh} and \gls*{ir} are separated.
	
	If the \gls*{eh} state is not known, the \gls*{pdf} of the input symbol amplitudes, $p_{r_x}$, which must be known at both \gls*{tx} and \gls*{ir}, has to be independent of $\xi$, i.e., $p^i_{r_x}(r) = p_{r_x}(r), \; \forall i \in \{1,2,...,S_\Xi\}$. 
	Then, the joint \gls*{pdf} $\boldsymbol{\pi}(r_x)$ and the average harvested power, $\overline{P}$, reduce as follows	
		\begin{equation}
		{\pi_i}(r_x) = {\gamma}_i \, p_{r_x}(r_x),
		\label{DistributionDecouplingEqn}
		\end{equation}
		\begin{equation}
		\overline{P}\big( \boldsymbol{\gamma}, p_{r_x} \big) = \sum_{i=1}^{S_\Xi} \int_{r_x} {\gamma}_i p_{r_x}(r_x) \tilde{P}_i(r_{x}) d r_x,
		\label{AverageRewardDecoupling}
		\end{equation}
	\noindent where vector $\boldsymbol{\gamma} = [\gamma_1, \gamma_2, ..., \gamma_{S_\Xi}]^\top$ collects the marginal probabilities of the \gls*{eh} states $\xi_i, i \in \{1,2,...,S_\Xi\}$.
	In order to obtain the boundary of the rate-power region, we formulate the following optimization problem:	
	
	\begin{subequations}
		\label{SeparateOptProblem}
		\begin{align}
		\label{SeparateOptProblem_Objective}
		\maximize_{\boldsymbol{\gamma}, p_{r_x} } \; & \overline{P}\big( \boldsymbol{\gamma}, p_{r_x} \big)	\quad \\
		\text{subject to}\quad &
		\label{SeparateOptProblem_Constr1}		
		{I}\big( p_{r_x} \big) \geq I_{\text{req}}, \\
		\label{SeparateOptProblem_Constr2}		
		& \int_{r_x} r_x^2 p_{r_x}(r_x) dr_x \leq \sigma_{r_x}^2, \\
		\label{SeparateOptProblem_Constr3}
		& |r_x| \leq r_x^\text{max}, \\	
		\label{SeparateOptProblem_Constr4}
		& \begin{aligned} \sum_{i=1}^{S_\Xi} \int_{r_x} {\gamma}_{i} p_{r_x}(r_x) \big( &\mathbf{1}_j(i) - \rho_{i,j}(r_x) \big) d r_x \\[-1.2ex] 
		&= 0, \; j \in \{1,2,...S_\Xi\},\end{aligned}
		\\
		\label{SeparateOptProblem_Constr5}
		& \int_{r_x} p_{r_x}(r_x) d r_x = 1,
		\\
		\label{SeparateOptProblem_Constr6}
		& \sum_{i=1}^{S_\Xi} {\gamma}_{i}= 1,
		\end{align}
	\end{subequations}
	\noindent where, similar to (\ref{MajorOptProblem}), we maximize the average harvested power at the \gls*{eh}, $\overline{P}$, subject to the minimum required mutual information, $I_\text{req}$, between \gls*{tx} and \gls*{ir}. 
	Note that in contrast to problem (\ref{MajorOptProblem}), since, in this case, the distribution $p_{r_x}$ is identical in each symbol interval, the expected mutual information $\overline{I} ( \boldsymbol{\pi} )$ is equal to the mutual information $I(p_{r_x})$ for any given state, cf. (\ref{SeparateOptProblem_Constr1}). 
	Constraints (\ref{SeparateOptProblem_Constr2}) and (\ref{SeparateOptProblem_Constr3}) ensure that the input distribution $p_{r_x}$ satisfies the \gls*{ap} and \gls*{pp} limits at the \gls*{tx}, respectively. 
	Furthermore, we have reformulated constraint (\ref{MajorOptProblem_Constr5}) as (\ref{SeparateOptProblem_Constr4}), whereas (\ref{MajorOptProblem_Constr4}) has been decomposed into (\ref{SeparateOptProblem_Constr5}) and (\ref{SeparateOptProblem_Constr6}).
	
	We observe that objective function (\ref{SeparateOptProblem_Objective}) and constraint (\ref{SeparateOptProblem_Constr4}) are not jointly concave and convex with respect to ${\boldsymbol{\gamma}}$ and $p_{r_x}$, respectively. 
	Hence, (\ref{SeparateOptProblem}) is a non-convex optimization problem, and therefore, the computation of its global optimal solution entails a high complexity.
	However, if we fix one of the variables, i.e., ${\boldsymbol{\gamma}}$ or $p_{r_x}$, both (\ref{SeparateOptProblem_Objective}) and (\ref{SeparateOptProblem_Constr4}) become linear in the other variable. 
	Hence, the subproblems obtained from (\ref{SeparateOptProblem}) by fixing either ${\boldsymbol{\gamma}}$ or $p_{r_x}$ are convex and can be solved efficiently.
	Therefore, in the following, to find a suboptimal solution of (\ref{SeparateOptProblem}), we adopt alternating optimization, e.g., \cite{Gorski2007}, which is known for its high efficiency and fast convergence speed. 
	The solution obtained by the proposed algorithm converges to a limit point of (\ref{SeparateOptProblem}).
	
	\subsection{Algorithm for Solving (\ref{SeparateOptProblem})}
	In the following, we develop an iterative algorithm, which involves an inner and an outer loop, to obtain a suboptimal solution of (\ref{SeparateOptProblem}). 
	In the outer loop of the algorithm, as in \cite{Si2014}, we relax the equality constraints in (\ref{SeparateOptProblem_Constr4}) to inequality constraints and tighten the relaxation in each iteration. 
	In the inner loop, adopting alternating optimization, we obtain a limit point for the relaxed version of problem (\ref{SeparateOptProblem}).
	
	\subsubsection{Outer Loop}
	We observe that the optimization variables in problem (\ref{SeparateOptProblem}) can be separated into two non-overlapping subsets, i.e., the \gls*{pdf} of symbol amplitudes, $p_{r_x}$, and the distribution of \gls*{eh} states, $\boldsymbol{\gamma}$, and therefore, alternating optimization is a promising approach for solving (\ref{SeparateOptProblem}) \cite{Grippo2000}.
	However, (\ref{SeparateOptProblem_Constr4}) imposes $S_\Xi$ equality constraints.
	Hence, applying alternating optimization directly to (\ref{SeparateOptProblem}) may lead to a strongly suboptimal solution since, in each iteration, the degrees of freedom for the optimization of $p_{r_x}$ and $\boldsymbol{\gamma}$ are very limited. 
	Thus, to overcome this issue, similar to \cite{Si2014}, in iteration $m$ of the outer loop, we relax the equality constraints in (\ref{SeparateOptProblem_Constr4}) to inequality constraints as follows 
		\begin{equation}
		\begin{aligned}	
			\Bigg\vert \int_{r_x} p_{r_x}(r_x) \sum_{i=1}^{S_\Xi} {\gamma}_{i} \big( \mathbf{1}_j(i) - \rho_{i,j}(r_x) & \big) d r_x \Bigg\vert \leq \epsilon^{\text{tol}}_m, \\[-0.5ex]
			&j \in \{1,2,...S_\Xi\}, 
		\end{aligned}	
		\label{RelaxedConstraint}
		\end{equation}	
	\noindent where $\epsilon^{\text{tol}}_m = \epsilon^{\text{tol}}_{m-1} \delta_{\epsilon}$ is the tolerance for the constraint violation, which will be tightened from one iteration of the outer loop to the next, and $\delta_{\epsilon} \in (0,1)$ is a constant factor. 
	
	Furthermore, we note that for a given distribution of symbol amplitudes $p_{r_x}$ which satisfies (\ref{SeparateOptProblem_Constr1}) - (\ref{SeparateOptProblem_Constr3}), (\ref{SeparateOptProblem_Constr5}), there exists a unique distribution of \gls*{eh} states $\boldsymbol{\gamma}$, such that the pair $\{p_{r_x}, \boldsymbol{\gamma}\}$ is in the feasible set of (\ref{SeparateOptProblem}) \cite{Norris1998}. 
	This distribution of \gls*{eh} states, $\boldsymbol{\gamma}$, can be obtained as the unique solution of the system of balance equations defined by constraints (\ref{SeparateOptProblem_Constr4}) and (\ref{SeparateOptProblem_Constr6}). 
	Moreover, we note that (\ref{SeparateOptProblem_Constr4}) and (\ref{SeparateOptProblem_Constr6}) can be rewritten in matrix form as ${\boldsymbol{R}(p_{r_x})} \boldsymbol{\gamma}  = \boldsymbol{e}$, where the elements of $\boldsymbol{R}(p_{r_x}) \in \mathbb{R}^{(S_\Xi+1) \times S_\Xi}$ are given by $R(p_{r_x})_{j,i} = \int_{r_x} p_{r_x}(r_x) \big( \mathbf{1}_j(i) - \rho_{i,j}(r_x) \big) d r_x$  and ${R}(p_{r_x})_{S_\Xi+1,i} = 1, i,j \in \{1,2,...,S_\Xi\}$, whereas the elements of $\boldsymbol{e} \in \mathbb{R}^{(S_\Xi+1)}$ are all equal to zero, i.e., ${e}_i = 0$ if $i~\in~\{1,2,...,S_\Xi\}$, except for the last element, which is ${e}_{S_\Xi+1} = 1$. 
	
	We note that one of the equations in (\ref{SeparateOptProblem_Constr4}) is redundant and, hence, the rank of matrix $\boldsymbol{R}(p_{r_x})$ is equal to $S_\Xi$ \cite{Norris1998}.
	Thus, for an initial feasible \gls*{pdf} of symbol amplitudes in iteration $m$, $p_{r_x}^{m,0}$, we obtain the corresponding distribution of states ${\boldsymbol{\gamma}}^{m,0}$ as follows
		\begin{equation}
		\boldsymbol{\gamma}^{m,0} = \big({\boldsymbol{R}(p_{r_x}^{m,0})}\big)^\dagger \boldsymbol{e} .
		\label{SubProblemFitStates}
		\end{equation}	
	Then, starting from the initial point $\{p_{r_x}^{m,0}, \boldsymbol{\gamma}^{m,0}\}$, we find a limit point of problem (\ref{SeparateOptProblem}) with constraint (\ref{SeparateOptProblem_Constr4}) relaxed to (\ref{RelaxedConstraint}), $\{p_{r_x}^{m,*}, \boldsymbol{\gamma}^{m,*}\}$, utilizing alternating optimization, which is implemented in the inner loop of the proposed algorithm. 
	Finally, we obtain the initial point for the next iteration $m+1$ from $\{p_{r_x}^{m,*}, \boldsymbol{\gamma}^{m,*}\}$ by setting $p_{r_x}^{m+1,0} = p_{r_x}^{m,*}$ and calculating the corresponding feasible distribution of \gls*{eh} states, $\boldsymbol{\gamma}^{m+1,0}$, as in (\ref{SubProblemFitStates}).
	
	\subsubsection{Inner Loop}	
	In the inner loop, exploiting alternating optimization, we solve the subproblem obtained in outer loop iteration $m$ by relaxing constraint (\ref{SeparateOptProblem_Constr4}) to (\ref{RelaxedConstraint}). To this end, we sequentially fix one of the optimization variables, i.e., $\boldsymbol{\gamma}$ or $p_{r_x}$, and solve the resulting convex subproblem with respect to the other variable.
	
	\textbf{Step 1}: 
	In the first step of the $n^\text{th}$ iteration of the inner loop, we optimize the \gls*{pdf} of the symbol amplitudes $p_{r_x}$ for the given distribution of states ${\boldsymbol{\gamma}}^{m,n-1}$ calculated in iteration $n-1$.
	Since constraint (\ref{SeparateOptProblem_Constr6}) does not depend on \gls*{pdf} $p_{r_x}$, in the current step, we obtain the \gls*{pdf} of symbol amplitudes $p_{r_x}^{m,n}$ as the solution of the following convex optimization subproblem 		
		\begin{align}
		\label{SubProblemActions}
		\maximize_{ p_{r_x} } \; & \overline{P}\big( {\boldsymbol{\gamma}}^{m,n-1}, p_{r_x} \big)  
		\quad \\
		\text{subject to}\quad & 
		\begin{aligned}\left| \int_{r_x} p_{r_x}(r_x) \sum_{i=1}^{S_\Xi} {\gamma}^{m,n-1}_{i} \big( \mathbf{1}_j(i)
		- \rho_{i,j}(r_x) \big) d r_x \right| \\ \leq \epsilon^{\text{tol}}_m, \; j \in \{1,2,...S_\Xi\} \end{aligned} \nonumber\\
		\nonumber
		&(\text{\ref{SeparateOptProblem_Constr1}}), (\text{\ref{SeparateOptProblem_Constr2}}), (\text{\ref{SeparateOptProblem_Constr3}}), (\text{\ref{SeparateOptProblem_Constr5}}).
		\end{align}	

	\textbf{Step 2}: 
	We note that constraints (\ref{SeparateOptProblem_Constr1}), (\ref{SeparateOptProblem_Constr2}), (\ref{SeparateOptProblem_Constr3}), and (\ref{SeparateOptProblem_Constr5}) do not depend on the distribution of states $\boldsymbol{\gamma}$. 
	Hence, for the given \gls*{pdf} $p_{r_x}^{m,n}$, we formulate the subproblem to obtain $\boldsymbol{\gamma}^{m,n}$ as follows 
		\begin{align}
		\label{SubProblemStates}
		\maximize_{ {\boldsymbol{\gamma}} } \; & \overline{P}\big( {\boldsymbol{\gamma}}, p_{r_x}^{m,n} \big)  
		\quad \\
		\text{subject to}\quad & 
		\begin{aligned}\left| \sum_{i=1}^{S_\Xi} {\gamma}_{i} \int_{r_x} p^{m,n}_{r_x}(r_x) \big( \mathbf{1}_j(i) - \rho_{i,j}(r_x) \big) d r_x \right|  \\ \leq \epsilon^{\text{tol}}_m, j \in \{1,2,...S_\Xi\}, \end{aligned} \nonumber	\\
		\nonumber
		&(\text{\ref{SeparateOptProblem_Constr6}}).
		\end{align}		
	The proposed optimization algorithm is summarized in {Algorithm~\ref{OptimizationAlgorithm}}.
		\begin{algorithm}[!t]				
			\SetAlgoNoLine%
			\SetKwFor{Foreach}{for each}{do}{end}		
			Initialize: Maximum number of iterations $M_{\text{max}}$, $N_{\text{max}}$, iteration indices $m=1$, $n=1$, initial tolerance $\epsilon^{\text{tol}}_1$, constant factor $\delta_\epsilon$, and initial distribution $p_{r_x}^{1,0}$ satisfying (\ref{SeparateOptProblem_Constr1})-(\ref{SeparateOptProblem_Constr3}), (\ref{SeparateOptProblem_Constr5}).	\\	
			\Repeat{\textbf{\upshape termination condition\footnotemark[5] is met or} $m=M_{\text{\upshape max}}+1$ }{		
				1.  For distribution $p_{r_x}^{m,0}$, set the elements of ${{\boldsymbol{R} (p_{r_x}^{m,0}) }}$ as ${R}(p_{r_x}^{m,0})_{i,j} = \int_{r_x} p_{r_x}^{m,0}(r_x) \big( \mathbf{1}_j(i) - \rho_{i,j}(r_x) \big) d r_x$ and ${R}(p_{r_x}^{m,0})_{i,S_\Xi+1} = 1$,  $i,j \in \{1,2,...,S_\Xi\}$\\
				2. Find the initial distribution of states $\boldsymbol{\gamma}^{m,0}$ from (\ref{SubProblemFitStates})\\									
				\Repeat{\textbf{\upshape termination condition\footnotemark[5] is met or} $n=N_{\text{\upshape max}}+1$ }{			
					a. For the given ${\boldsymbol{\gamma}}^{m,n-1}$, solve convex problem (\ref{SubProblemActions}) with CVX \cite{Grant2015}, and store the intermediate \gls*{pdf} of transmit symbol amplitudes $p_{r_x}^{m,n}$\\
					b. For the given $p_{r_x}^{m,n}$, solve convex problem (\ref{SubProblemStates}) with CVX \cite{Grant2015}, and store the intermediate distribution of states ${\boldsymbol{\gamma}}^{m,n}$\\
					c. Set $n = n+1$\\
				}				
				3. Set initial value for the next iteration $p_{r_x}^{m+1,0} = p_{r_x}^{m,n-1}$\\
				4. Update the constraint violation tolerance $\epsilon^{\text{tol}}_{m+1} = \epsilon^{\text{tol}}_{m} \delta_{\epsilon}$\\
				5. Set $m = m+1$\\ 
			}				
			\textbf{Output:} 
			$ {\boldsymbol{\gamma}}^{M_\text{max}, N_\text{max}}, p_{r_x}^{M_\text{max}, N_\text{max}} $
			\caption{\strut Iterative algorithm for solving optimization problem (\ref{SeparateOptProblem})}
			\label{OptimizationAlgorithm}
		\end{algorithm}	

	{In the following, we discuss the convergence of Algorithm~\ref{OptimizationAlgorithm}.}
	First, we note that problem (\ref{SubProblemActions}) is convex, whereas (\ref{SubProblemStates}) is linear and, hence, both problems can be efficiently solved using standard numerical optimization tools, such as CVX.	
	{As shown in \cite{Gorski2007}, starting from the feasible point $\{ {\boldsymbol{\gamma}}^{m,0} , p_{r_x}^{m,0} \}$, in the inner loop, the sequence $\{ {\boldsymbol{\gamma}}^{m,n} , p_{r_x}^{m,n} \}$ converges monotonically to a limit point of the corresponding relaxed subproblem, $\{ {\boldsymbol{\gamma}}^{m,*} , p_{r_x}^{m,*} \}$. 
		\footnotetext[5]{Here, in order to verify convergence, as in \cite{Nocedal2006}, a termination condition can be used in the inner and outer loops of the algorithm, e.g., $\vert p_{r_x}^{m,n-1}~-~p_{r_x}^{m,n-2} \vert \leq \epsilon^i_p$ and $\vert p_{r_x}^{m,0} - p_{r_x}^{m-1,0} \vert \leq \epsilon^o_p$, respectively, where $\vert \cdot \vert$ is the $\text{L}_1$-norm, and $\epsilon^i_p$ and $\epsilon^o_p$ are pre-defined maximum values for the errors in the inner and outer loops, respectively.\label{footnote:termination}}
	Moreover, as outer loop iteration $m$ increases, the sequence of feasible sets of the relaxed subproblems determined by constraints (\ref{SeparateOptProblem_Constr1}) - (\ref{SeparateOptProblem_Constr3}), (\ref{SeparateOptProblem_Constr5}), (\ref{SeparateOptProblem_Constr6}), and (\ref{RelaxedConstraint}) converges to the feasible set of the initial problem (\ref{SeparateOptProblem}). 
	Hence, the sequence of feasible points $\{ {\boldsymbol{\gamma}}^{m, 0}, p_{r_x}^{m, 0} \}$ converges to a limit point of (\ref{SeparateOptProblem}) denoted by $\{ {\boldsymbol{\gamma}}^{*}, p_{r_x}^{*} \}$.
	Furthermore, since the monotonicity of the sequence $\overline{P}\big( {\boldsymbol{\gamma}}^{m,0}, p_{r_x}^{m,0} \big), m \in \{1,2,..\},$ cannot be guaranteed in general, as a suboptimal solution of (\ref{SeparateOptProblem}), one may choose the pair $\langle {\boldsymbol{\gamma}}', p_{r_x}' \rangle = \argmax_{ \langle\boldsymbol{\gamma}, p_{r_x}\rangle \in \mathcal{G} } \overline{P}\big( \boldsymbol{\gamma}, p_{r_x} \big)$, where $\mathcal{G}~=~\big\{ \langle\boldsymbol{\gamma}^{m, 0}, p_{r_x}^{m, 0}\rangle \mid m \in \{1,2,...\} \big\}$ is the set of feasible points obtained in the outer loop.
	However, we observed monotonic convergence of the sequence $\overline{P}(\boldsymbol{\gamma}^{m, 0}, p_{r_x}^{m, 0})$ in our simulations. Therefore, as a solution of (\ref{SeparateOptProblem}), we adopt the pair $\{ {\boldsymbol{\gamma}}^{M_\text{max}, 0}, p_{r_x}^{M_\text{max}, 0} \}$, which is a limit point of (\ref{SeparateOptProblem}) provided that the maximum number of iterations of the outer loop, $M_\text{max}$, is large enough to ensure that $\epsilon^{\text{tol}}_{M_\text{max}} \approx 0$. }	
	
	{\textbf{Remark 1:~} The computational complexity of Algorithm 1 depends on the methods used to solve subproblems (\ref{SubProblemActions}) and (\ref{SubProblemStates}). 
	The computational complexity order of a widely used interior-point algorithm \cite{Boyd2004} for
	linear and conic problems with $n$ variables and $m$ constraints is $\mathcal{O}\Big( n^3 + n^2 m  \Big)$, where $\mathcal{O}\big(\cdot\big)$ is the big-O notation \cite{Nesterov1994, Nemirovski2004, AkleSerrano2015}. 
	Therefore, the computational complexity of the proposed algorithm per iteration of the inner loop as a function of the size of the constellation set\footnotemark[6] $S$ of the transmit symbols and the MDP state space size $S_\Xi$ is given by $\mathcal{O}\big(S_\Xi^3 + S_\Xi S^2 + S^3\big)$.
	Note that algorithms with polynomial time complexity are usually considered to be fast algorithms in the literature \cite{Nemirovski2004, Nocedal2006}.
	Since the EH circuit parameters are known at the TX, optimization problem (\ref{SeparateOptProblem}) can be efficiently solved at TX with Algorithm 1.
	}	
	\footnotetext[6]{ To determine the pdf of the transmit symbol amplitudes, $p_{r_x}$, numerically, as in, e.g., \cite{Morsi2019, Smith1971}, in our simulations, we assume that the symbol amplitudes $r_x$ are taken from a constellation set of size $S$, see Section \ref{SectionSimulationParameters}.}
	\begin{table*}[!ht]
		\normalsize
		\centering
		\caption{EH Circuit Parameters.}
		\label{TableCircuits}
		\def\arraystretch{1.2}
		\begin{tabular}{|l|c|c|c|c|} 
			\hline
			Rectifier circuit & \multicolumn{2}{c|}{Half-wave rectifier} & \multicolumn{2}{c|}{Full-wave rectifier}\\
			\hline
			Input power level for MC design & $\SI{-13}{\deci\belmilliwatt}$ & $\SI{0}{\deci\belmilliwatt}$ & $\SI{-13}{\deci\belmilliwatt}$ & $\SI{0}{\deci\belmilliwatt}$ \\
			\hline
			\hline
			Diode model & \multicolumn{4}{c|}{SMS7630}\\
			\hline
			Antenna resistance  & \multicolumn{4}{c|}{$R_s = \SI{50}{\ohm}$}\\
			\hline
			Load capacitor & \multicolumn{4}{c|}{$C_L = \SI{1}{\nano\farad}$}\\
			\hline
			Load resistor & \multicolumn{4}{c|}{$R_L = \SI{10}{\kilo\ohm}$}\\
			\hline
			Inductance of the MC & $L_1 = \SI{26.7}{\nano\henry}$ & $L_1 = \SI{9.62}{\nano\henry}$ & $L_1 = \SI{23.2}{\nano\henry}$ & $L_1 = \SI{11.1}{\nano\henry}$\\
			\hline
			\multirow{2}{*}{Capacitance of the MC} & $C_1 = \SI{0.73}{\pico\farad}$ & $C_1 = \SI{1.41}{\pico\farad}$ & $C_1 = \SI{0.3}{\pico\farad}$ & $C_1 = \SI{2.72}{\pico\farad}$\\
			&- & $C_2 = \SI{0.375}{\pico\farad}$ & -  & $C_2 = \SI{0.3}{\pico\farad}$ \\
			\hline
			{\SI{3}{\dB} bandwidth of the MC} & \SI{270}{\mega\hertz} & \SI{280}{\mega\hertz} & \SI{310}{\mega\hertz}& \SI{300}{\mega\hertz} \\
			\hline
		\end{tabular}
	\end{table*}
	\subsection{Infinitely Large Symbol Duration}
	\label{SectionInfiniteDuration}	
	\begin{figure}[t]
		\centering
		\includegraphics[width=0.49\textwidth]{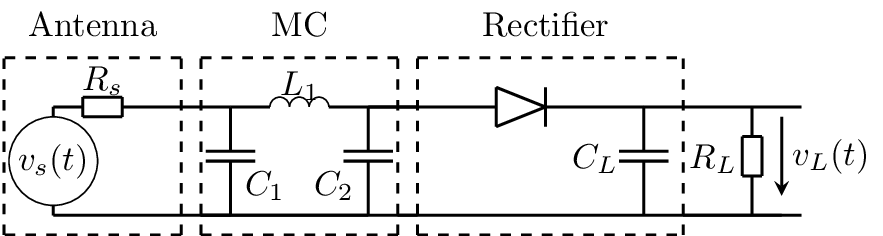}				
		\caption{\gls*{eh} circuit model comprising an MC, a single diode half-wave rectifier, and a capacitor $C_L$ as part of an \gls*{lp}.}
		\label{EHSingleDiodeRectifier_Fig}	
	\end{figure}
	In the following, we consider the special case of a SWIPT system with symbol duration $T \to \infty$. First, in this case, we observe that the reactive element of the \gls*{lp} in the \gls*{eh} circuit saturates to a voltage level which depends on the received symbol only \cite{Horowitz1989}. 
	Hence, the next state $\xi[k]$ of the \gls*{mdp} does not depend on the previous state $\xi[k-1]$, i.e., $\mathrm{Pr} \{ \xi[k] \mid r_x[k], \xi[k~-~1]\} = \mathrm{Pr} \{ \xi[k] \mid r_x[k]\}$. 
	Then, the transition \gls*{pdf} of this process reduces to $\rho_{i,j}(r_{x}) = \rho_i(r_{x}),$ $\forall~j~\in~\{1,2,...,S_\Xi\}$. 
	Moreover, in this scenario, the power harvested by the \gls*{eh} before the voltage level at the reactive element in the \gls*{lp} saturates becomes negligible as $T \to \infty$. 
	Hence, the amount of power harvested by the \gls*{eh} during time slot $k$ does not depend on the initial state $\xi[k]$, i.e., $\tilde{P}_i(r_{x}) = P'(\tilde{v}_i, |h_E| r_{x})~=~P'(|h_E| r_{x}), \forall i~\in~\{1,2,...,S_\Xi\}$. 
	Thus, the average harvested power in (\ref{ExpandedAverageRewardEqn}) depends on the \gls*{pdf} $p_{r_x}$ only, i.e., 	$\overline{P}( \boldsymbol{\pi} ) = \overline{P}\big( p_{r_x} \big) = \mathbb{E}_{r_{x}} \big\{ {P}'(|h_E| r_{x}) \big\}$. 
	In this case, the average harvested power $\overline{P}\big( p_{r_x} \big) = \int_{r_x} p_{r_x}(r_x) {P}'(|h_E| r_{x}) \, d r_x $ is maximized by optimizing the \gls*{pdf} of the transmitted symbols $p_{r_x}$ which is independent of $\xi$, and hence, is identical in each symbol interval.	
\setcounter{footnote}{5}	
	Consequently, (\ref{MajorOptProblem}) simplifies as follows
	\begin{subequations}
		\label{InifiniteTOptProblem}
		\begin{align}
		\label{InifiniteTOptProblem_Objective}
		\maximize_{ p_{r_x} } \; & \overline{P}\big( p_{r_x} \big) \quad \\
		\text{subject to}\quad &
		\label{InifiniteTOptProblem_Constr1}		
		I\big( p_{r_x} \big) \geq I_{\text{req}}, \\		
		\label{InifiniteTOptProblem_Constr2}
		& \int_{r_x} p_{r_x}(r_x) \, dr_x  = 1, \\
		& (\text{\ref{MajorOptProblem_Constr2}}), (\text{\ref{MajorOptProblem_Constr3}}).
		\end{align}
	\end{subequations}	
			
	We observe that optimization problem (\ref{InifiniteTOptProblem}) is convex and can be solved using standard numerical solvers, such as CVX. 
	Moreover, problem (\ref{InifiniteTOptProblem}) is equivalent to \cite[Eq. (18)]{Morsi2019}, where an analytical expression for $P'(r_E)$ was derived assuming a half-wave rectifier circuit with a single diode at the \gls*{eh}, a clipping model for the \gls*{eh} saturation, and perfect impedance matching. 
	Thus, for this special case, (\ref{InifiniteTOptProblem}), and hence, (\ref{MajorOptProblem}) yield the same solution as the one obtained in \cite{Morsi2019}.
	We note, however, that in this paper, we use a learning based model for $P'(r_E)$, which is applicable for imperfect impedance matching and arbitrary rectifier circuits.	
	\begin{table*}
		\normalsize
		\begin{center}
			\caption{Simulation Parameters.}
			\label{TableSystem}
			\def\arraystretch{1.4}
			\begin{tabular}{|l|c||l|c|} 
				\hline 
				Carrier frequency & $f_c = 2.45 \, \text{GHz}$ & \gls*{awgn} variance at \gls*{ir} & $\sigma_n^2 = \SI{-70}{\deci\belmilliwatt}$ \\
				\hline
				\gls*{pp} limit at \gls*{tx} & $P^{\text{\gls*{tx}}}_{\text{max}} = \SI{50}{\dBm}$ & \gls*{ap} limit at \gls*{tx} & $\sigma_x^2 = \SI{42}{\deci\belmilliwatt}$ \\
				\hline
				Constellation size & $S = 64$ & \gls*{eh} state space size & $S_\Xi = 50$ \\
				\hline				
				Pathloss exponent of \gls*{ir} channel & $\alpha_{I} = 3$ & Pathloss exponent of \gls*{eh} channel & $\alpha_{E} = 2$ \\
				\hline
				\multirow{3}{*}{Distance between \gls*{tx} and \gls*{ir} } & \multirow{3}{*}{ $d_{I} = \SI{40}{\meter}$ } & \multirow{3}{*}{Distance between \gls*{tx} and \gls*{eh}} & LP regime: $d_{E} = \SI{20}{\meter}$ \\ & & & MP regime: $d_{E} = \SI{10}{\meter}$ \\ & & & HP regime: $d_{E} = \SI{2}{\meter}$ \\
				\hline				
				Initial tolerance in Algorithm 1 & $\epsilon^\text{tol}_1 = 0.5$ & Tolerance decrease factor in Algorithm 1 & $\delta_\epsilon = 0.5$ \\
				\hline
				{Maximum number of iterations} & $M_{\text{max}} = 15$ & Maximum error tolerance in the & $\epsilon_{p}^i = 10^{-7}$  \\
				of Algorithm 1	& $N_{\text{max}} = 10$ &termination conditions in Algorithm 1 & $\epsilon_{p}^o = 10^{-7}$ \\ 
				\hline
			\end{tabular}
		\end{center}
	\end{table*}	
	\subsection{Learning Based Model For \gls*{eh} Circuits}
	\label{SectionNeuralNetwork}	
	In this section, we discuss a learning based approach to approximate functions $f_v(v, r_{E})$ and $P'(v, r_{E})$, which are required for calculation of the transition \gls*{pdf} $\rho_{i,j}(r_x)$ and the harvested power $\tilde{P}_i(r_{x})$ in optimization problems (\ref{MajorOptProblem}), (\ref{SeparateOptProblem}), and (\ref{InifiniteTOptProblem}).
	As discussed in Section \ref{SectionEnergyHarvester}, for practical \gls*{eh}s, given load voltage level, $v$, and the amplitude of the received symbol, $r_E = |h_E|r_x$, it is not tractable to develop exact analytical expressions of the transition function $f_v(v, r_{E})$ and the reward associated with the corresponding transition $P'(\tilde{v}, r_{E})$, where $\tilde{v}$ is the quantized voltage level $v$, because of the imperfections of the \gls*{eh} circuit. 
	However, due to the universal approximation theorem for DNNs \cite{Hanin2017, Hornik1989}, it is possible to estimate the values of these functions with DNNs.  
	\begin{figure}[t]
		\centering
		\includegraphics[width=0.48\textwidth]{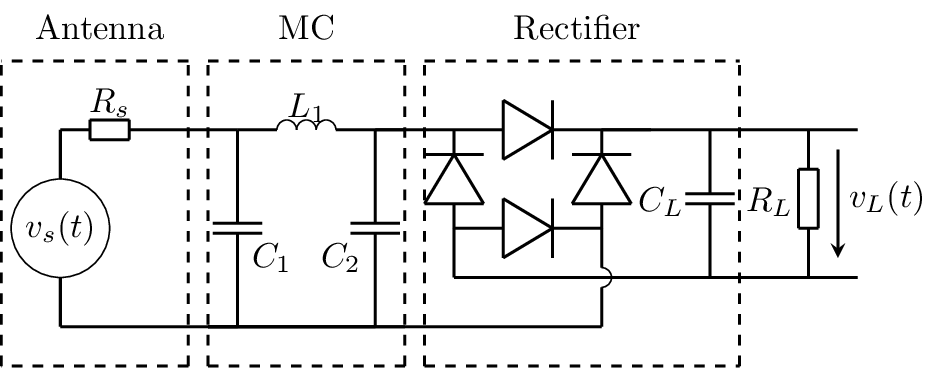}
		\caption{\gls*{eh} circuit model comprising an MC, a full-wave rectifier based on the bridge diode configuration, and a capacitor $C_L$ as part of an \gls*{lp}.}
		\label{EH_Bridge_Fig}	
	\end{figure}	
	
	A DNN is comprised of nodes organized into groups called layers \cite{Goodfellow2016}. 
	The outputs of the nodes in layer $i$ are collected in vector $\boldsymbol{h}^{(i)} \in \mathbb{R}^{\omega_i}$, where $\omega_i$ is the number of nodes in layer $i$. The output vector of the first layer is given by $\boldsymbol{h}^{(1)} = \boldsymbol{g}^{(1)} (\boldsymbol{W}^{(1)} \boldsymbol{x} + \boldsymbol{b}^{(1)})$, where $\boldsymbol{x}$ is the input vector of the DNN, whereas the outputs of the hidden layers, i.e., the layers following the first layer, are functions of the outputs of the layers preceding them, i.e., $\boldsymbol{h}^{(i)} = \boldsymbol{g}^{(i)} (\boldsymbol{W}^{(i)} \boldsymbol{h}^{(i-1)} + \boldsymbol{b}^{(i)}), i \in \{2, ..., N_l\}$, where $\boldsymbol{g}^{(i)}(\cdot)$, $\boldsymbol{W}^{(i)} \in \mathbb{R}^{\omega_i \times \omega_{i-1}}$, and $\boldsymbol{b}^{(i)} \in \mathbb{R}^{\omega_i}$ are the activation function, the weight matrix, and the bias vector adopted in layer $i$, respectively, and $N_l$ is the total number of layers of the DNN \cite{Goodfellow2016}.
	Commonly used activation functions for DNNs, $\boldsymbol{g}^{(i)}(\cdot)$, include ReLU and sigmoid functions \cite{Goodfellow2016}, which are computed element-wise.
	The parameters of the DNN, i.e., the weight matrices and bias vectors of each layer, are collected in a set $\boldsymbol{\Omega} = \{\boldsymbol{W}^{(1)}, \boldsymbol{b}^{(1)}, \boldsymbol{W}^{(2)}, \boldsymbol{b}^{(2)}, ..., \boldsymbol{W}^{(N_l)}, \boldsymbol{b}^{(N_l)}\}$ and can be obtained using a gradient-based back-propagation algorithm \cite{Goodfellow2016}.
	The last layer of a DNN is usually composed of a single node, i.e., $\omega_{N_l} = 1$, whose output is the output of the DNN, i.e., ${h}^{(N_l)} = \mathcal{N}(\boldsymbol{x}, \boldsymbol{\Omega})$. 
	Finally, we note that as shown in \cite{Hanin2017, Hornik1989}, a DNN with at least one hidden layer can approximate any Borel measurable function with any desired non-zero error.
	  	
	Here, we train two DNNs, $\hat{f}_{v}(v, r_E) = \mathcal{N}_1(v, r_E, \boldsymbol{\Omega}_1)$ and $\hat{P}'(v, r_{E}) = \mathcal{N}_2(v, r_{E}, \boldsymbol{\Omega}_2)$, where $\boldsymbol{x} = [v, r_E]^\top$ is the input vector of the \gls*{dnn}s, and $\hat{f}_{v}(v, r_E)$ and $\hat{P}'(v, r_{E})$ are the approximations of ${f}_{v}(v, r_E)$ and ${P}'(v, r_{E})$, respectively. 	
	Since the related approximation error depends on the network size \cite{Hanin2017}, the numbers of nodes of DNNs $\mathcal{N}_1$ and $\mathcal{N}_2$ have to be chosen properly.
	For our simulations, we use DNNs where the hidden nodes employ the ReLU activation functions. 
	In the output layer, as the suitability of the logistic function for modeling saturation effects in \gls*{eh} circuits was demonstrated in \cite{Boshkovska2015}, we use the sigmoid activation function.	
	
	In this paper, we consider two different rectenna \gls*{eh} circuits, namely, a single diode half-wave rectifier \cite{Morsi2019}, cf. Fig. \ref{EHSingleDiodeRectifier_Fig}, and a bridge full-wave rectifier \cite{Tietze2012}, cf. Fig. \ref{EH_Bridge_Fig}.
	For both rectenna circuits, we designed two different LC matching networks, fine-tuned for input signal frequency $2.45 \, \text{GHz}$ and input power values $\SI{-13}{\deci\belmilliwatt}$ and $\SI{0}{\deci\belmilliwatt}$ representing low and high input power levels, respectively. 
	As discussed in Section \ref{SectionEnergyHarvester}, although the MCs also include reactive elements, the memory introduced by the MC can be neglected in the \gls*{mdp} model since the MCs behave as band-pass filters, whose bandwidths are much larger than the considered symbol rates $\frac{1}{T}$, which do not exceed $\SI{200}{\kilo\hertz}$ in our simulations.
	The adopted \gls*{eh} circuit parameters are specified in Table \ref{TableCircuits}.
	
	Data for the training of the DNNs can be obtained from a circuit simulator, such as ADS \cite{ADS2017}. 
	To train the DNNs, we randomly generate the \gls*{iid} amplitudes of the received symbols, $r_{E}$, that are uniformly distributed over a space of symbols that can be realistically received by the \gls*{eh} and determine the corresponding 4-tuples $\big\{ {P}'(v_\nu, r_E), f_v(v_\nu, r_{E}), v_\nu, r_E  \big\}$ using the circuit simulator. 
	Specifically, we used 11000, 3000, and 750 4-tuples for training, validation, and testing of the \gls*{dnn}s for all considered circuits, respectively. For training, we used the Adam optimization algorithm \cite{Kingma2014} and the mean absolute percentage loss function, see e.g. \cite{Myttenaere2016}. 
	{We note that the parameters of the DNNs $\mathcal{N}_1(v, r_{E}, \boldsymbol{\Omega}_1)$ and  $\mathcal{N}_2(v, r_{E}, \boldsymbol{\Omega}_2)$ depend on the modeled EH circuit but not on the EH channel gain. Thus, the DNNs can be pre-trained offline at a high performance computing node.
	} 
	
	To minimize the estimation error measured on the test set for a given training set size, we trained several \gls*{dnn}s with different numbers of layers to find the best setting. 
	We found that for all considered rectenna circuits, the values of the mean absolute percentage error for DNNs $\mathcal{N}_1(v, r_{E}, \boldsymbol{\Omega}_1)$ and $\mathcal{N}_2(v, r_E, \boldsymbol{\Omega}_2)$ do not significantly decrease if the size of the DNNs increases beyond 7 layers and 15 nodes per hidden layer. 
	The parameters of each network obtained after training are saved to be used for estimation of the transition \gls*{pdf} $\rho_{i,j}(r_x)$ and the power values $\tilde{P}_i(r_{x})$, respectively, as needed for solving optimization problems (\ref{MajorOptProblem}), (\ref{SeparateOptProblem}), and (\ref{InifiniteTOptProblem}).
			
	\section{Simulation Results}
	\label{SectionNumericalResults}	
	\begin{figure}[!t]
		\centering
		\includegraphics[width=0.5\textwidth]{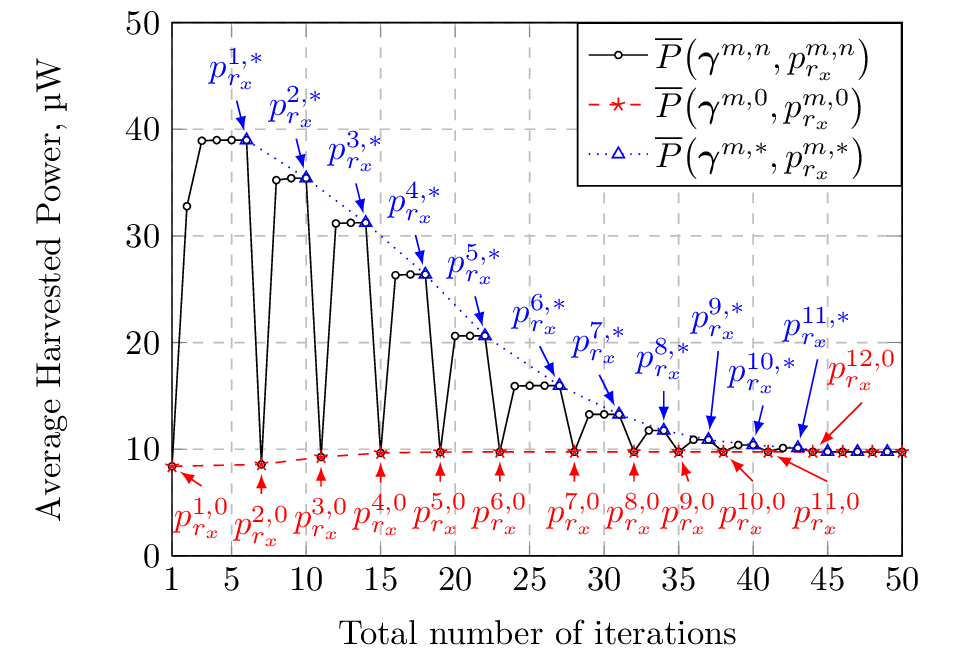}
		\caption{Convergence of Algorithm 1.}
		\label{Fig:Convergence}
	\end{figure}
	\begin{figure*}[!ht]	
	\centering	 	
	\subfigure[Rate-power regions for $T = \SI{100}{\micro\second}$]{%
		\includegraphics[width=.48\textwidth]{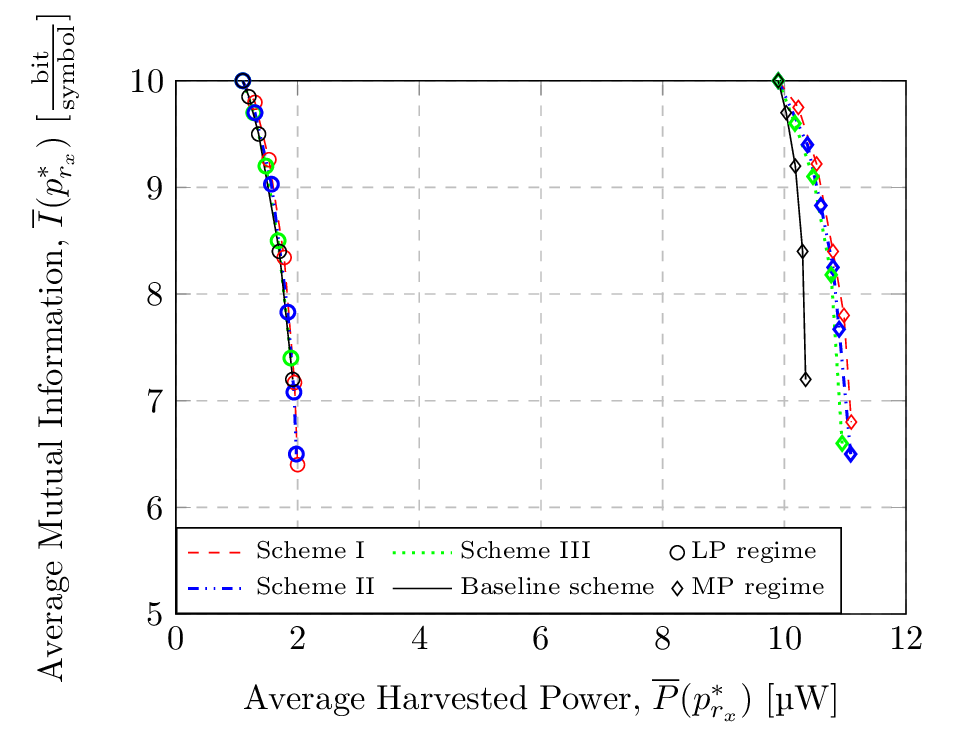} \label{fig:Val100} 
	}  \quad
	\subfigure[Rate-power regions for the \gls*{mpr} regime]{%
		\includegraphics[width=.48\textwidth]{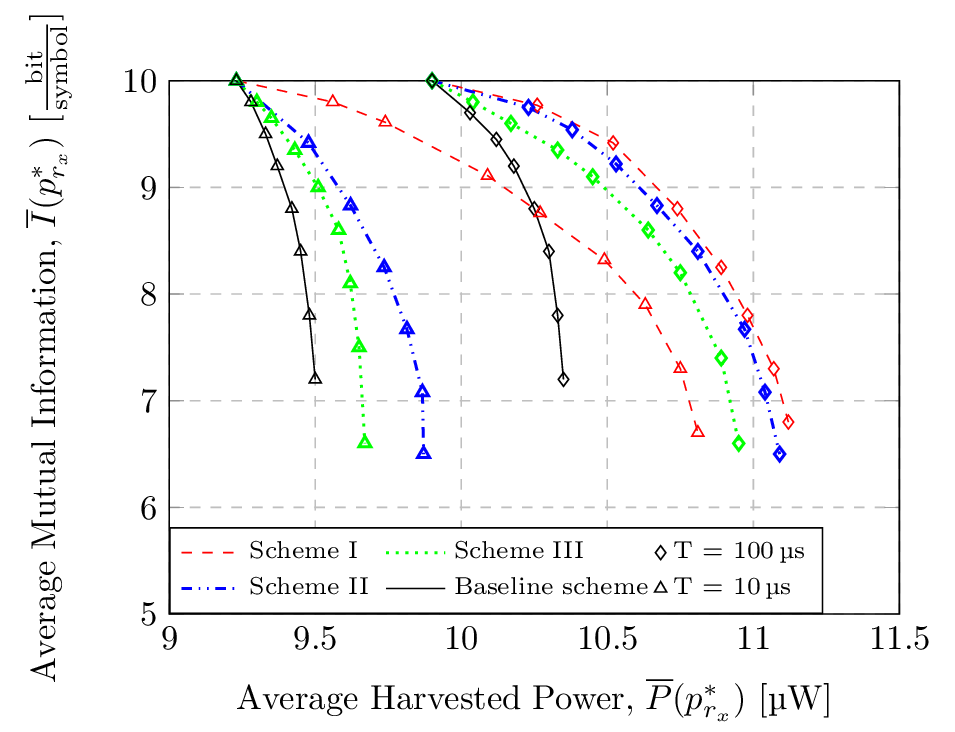} \label{fig:ValMP} 
	} 
	
	\caption{Rate-power region for Schemes I, II, and III, and for the baseline scheme.}
	\label{Validation_Fig}
	\centering
	\end{figure*}		
	In this section, for convenience, we refer to the operational modes corresponding to the solutions of problems (\ref{MajorOptProblem}), (\ref{SeparateOptProblem}), and (\ref{InifiniteTOptProblem}) as Scheme I, Scheme II, and Scheme III, respectively.
	In the following, after specifying the simulation parameters, we first study {the convergence of Algorithm 1}. Then, we validate the proposed model by comparing the rate-power regions of the \gls*{swipt} system obtained for Schemes I-III with that obtained for the scheme proposed in \cite{Morsi2019}. Subsequently, we investigate the impact of impedance mismatch between antenna and rectifier on the rate-power region. 
	Finally, we study the dependence of the rate-power region on the symbol duration $T$, the type of \gls*{eh} circuit, and the \gls*{eh} input power level, respectively.
	\subsection{Simulation Parameters}
	\label{SectionSimulationParameters}			
	For all simulations, we adopted uniformly spaced symbol amplitudes $r_x$, i.e., $r_{x_k}~=~\frac{k}{S-1} r_x^{\text{max}}$, $k = {0,1,...,S-1}$, with maximum symbol amplitude $r_x^{\text{max}}~=~10^{{P^{\text{\gls*{tx}}}_{\text{max}}}/{20}}$, where $P^{\text{\gls*{tx}}}_{\text{max}}$ is the \gls*{pp} limit at the \gls*{tx} in $\SI{}{\dBm}$ and $S$ is the constellation size. 	
	{As in \cite{Goldsmith2005}, the IR and EH channel gains are modeled as $\vert h_l \vert^2 = \vert \tilde{h}_l \vert^2 \Big( \frac{c }{4 \pi f_c d_0} \Big)^2 \Big(\frac{d_0}{d_l}\Big)^{\alpha_l}$, $l \in \{I,E\}$, where $\vert \tilde{h}_l \vert$ is the small scale fading coefficient, which is kept constant and equal to $1$ unless specified otherwise, $c$ denotes the speed of light, $\alpha_l$ is the pathloss exponent, $d_0 \leq d_l$ is the reference distance, which is set to $d_0 = \SI{1}{\meter}$ in our simulations, and $d_I$ and $d_E$ represent the distances between TX and IR and between TX and EH, respectively.} 
	Moreover, we consider three different input power regimes at the \gls*{eh}, namely, the \gls*{spr}, the \gls*{mpr}, and the \gls*{lpr} regimes characterized by different distances $d_{E}$ between \gls*{tx} and \gls*{eh}. 
	The adopted simulation parameters are summarized in Table~\ref{TableSystem}.	
	 {\subsection{Convergence of Algorithm 1}	
	 	\label{SectionConvergence}	
 	In Fig. \ref{Fig:Convergence}, we investigate the convergence of Algorithm 1 for the half-wave rectifier, an MC fine-tuned for $\SI{-13}{\dBm}$, a symbol duration of $T = \SI{10}{\micro\second}$, the MP regime for the EH, and a required mutual information between the TX and IR of $I_\text{req}~=~\SI{7.8}{\frac{\bit}{symbol}}$. For the termination conditions in Algorithm 1, we adopt error tolerance values of $\epsilon_{p}^i = 10^{-7}$ and $\epsilon_{p}^o = 10^{-7}$ for the inner and outer loops, respectively, see Footnote~\ref{footnote:termination}.
 	For all simulation parameters, the values in Table~\ref{TableSystem} are adopted. 
 	The maximum numbers of iterations, $N_\text{max}$ and $M_\text{max}$, are chosen sufficiently large to ensure that the termination conditions for the inner and outer loops of Algorithm 1 are met, respectively. 	  		
 	
 	In Fig. \ref{Fig:Convergence}, we observe that in each iteration of the outer loop $m$, starting from a feasible point of the initial problem $\{ {\boldsymbol{\gamma}}^{m,0}, p_{r_x}^{m,0} \}$, the proposed algorithm monotonically converges to a limit point of the corresponding relaxed subproblem $\{ {\boldsymbol{\gamma}}^{m,*}, p_{r_x}^{m,*} \}$. 
 	Also, we note that in the first iteration of the outer loop, the limit point of the relaxed subproblem, $\{ {\boldsymbol{\gamma}}^{m,*}, p_{r_x}^{m,*} \}$ yields a significantly higher average harvested power than the corresponding initial feasible point for the next iteration $\{ {\boldsymbol{\gamma}}^{m+1,0}, p_{r_x}^{m+1,0} \}$, i.e., $\overline{P}({\boldsymbol{\gamma}}^{m,*}, p_{r_x}^{m,*}) > \overline{P}({\boldsymbol{\gamma}}^{m+1,0}, p_{r_x}^{m,*}) = \overline{P}({\boldsymbol{\gamma}}^{m+1,0}, p_{r_x}^{m+1,0})$.
 	However, as the outer loop iterations $m$ increase, both the sequence of limit points $\{ {\boldsymbol{\gamma}}^{m,*}, p_{r_x}^{m,*} \}$ and the sequence of feasible points $\{ {\boldsymbol{\gamma}}^{m,0}, p_{r_x}^{m,0} \}$ monotonically converge to a limit point of (\ref{SeparateOptProblem}) within the specified error tolerance.
 	 
 	Exhaustive simulations have shown that for convergence to a limit point, i.e., to satisfy the termination conditions, Algorithm 1 requires at most $N_\text{max} = 5$ and $M_\text{max} = 12$ iterations in the inner and outer loops, respectively. 
 	Therefore, in the following, for the inner and outer loops, we set the maximum numbers of iterations to $N_\text{max} = 10$ and $M_\text{max} = 15$, respectively, cf. Table \ref{TableSystem}. } 
	\subsection{Model Validation}
	\label{SectionModelValidation}	
	\begin{figure*}[!t]	
		\centering		
		\subfigure[Half-wave rectifier]{%
			\includegraphics[width=.48\textwidth]{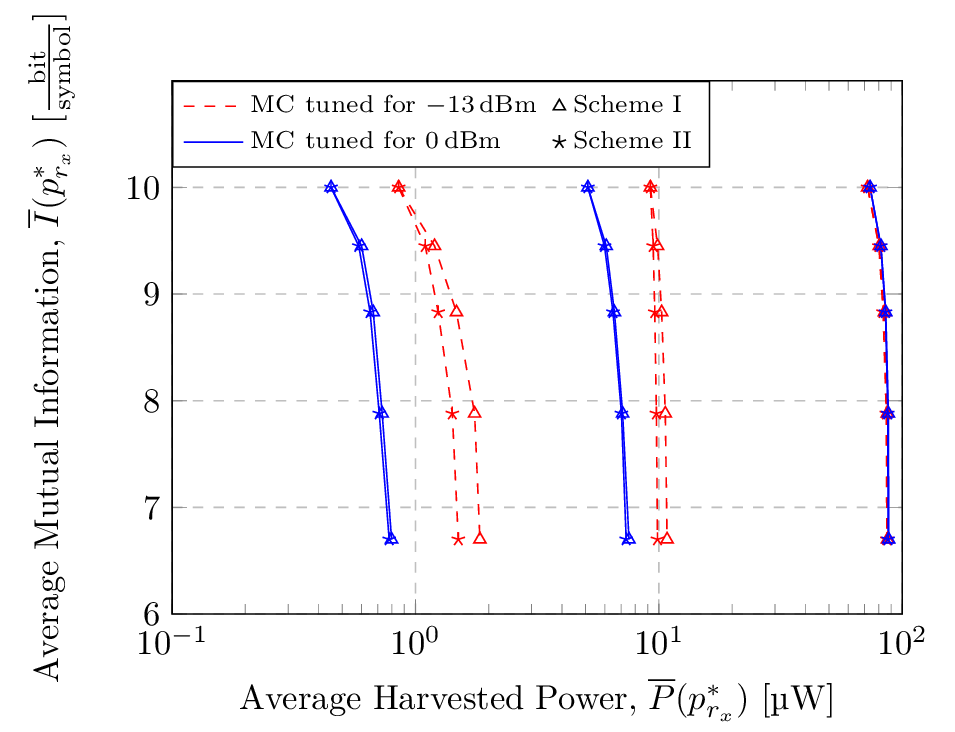} \label{fig:MatchingHWR} 
		} \quad
		\subfigure[Full-wave rectifier]{%
			\includegraphics[width=.48\textwidth]{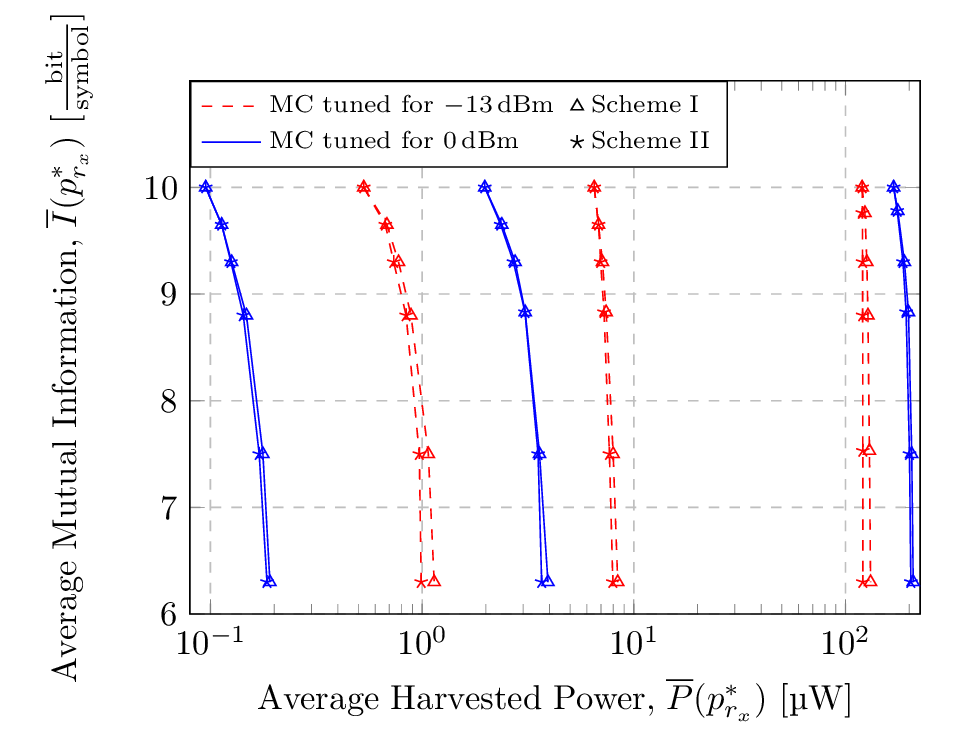} \label{fig:MatchingFWR} 
		} 
		\caption{Rate-power region for Schemes I and II for $T = \SI{10}{\micro\second}$.}
		\label{Mismatch_Fig}
		\centering
	\end{figure*}	
	\begin{figure*}
		\centering
		
		\subfigure[Half-wave rectifier]{%
			\includegraphics[width=.48\textwidth]{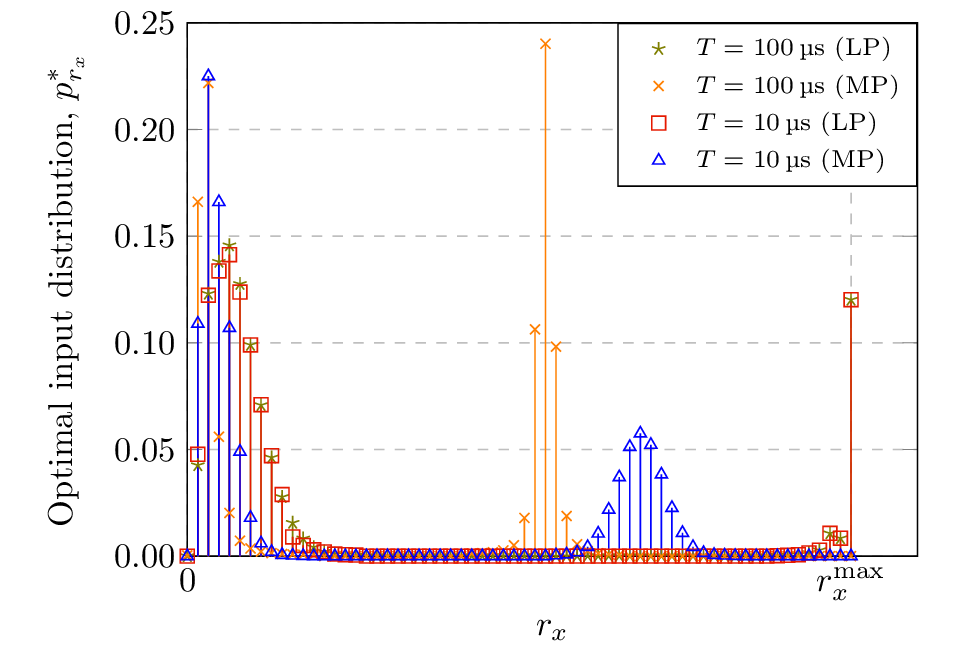} \label{fig::DistrHWR} 
		} \quad
		\subfigure[Full-wave rectifier]{%
			\includegraphics[width=.48\textwidth]{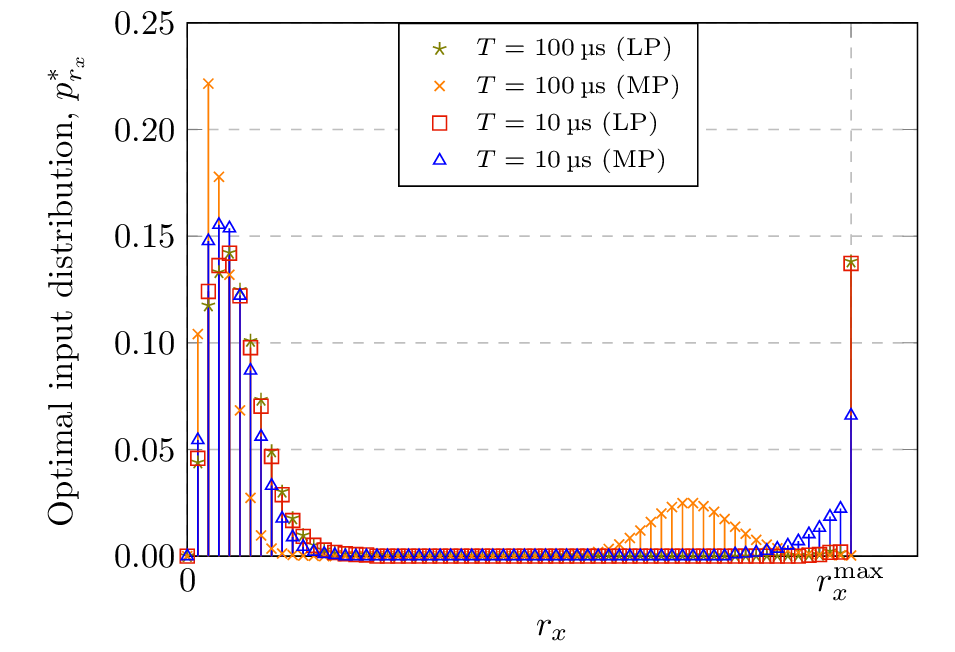} \label{fig::DistrFWR} 
		}

		\caption{Optimal input distribution for Scheme II with $I_{\text{req}} = \SI{6.5}{\frac{\bit}{symbol}}$.}
		\label{Distributions_Fig}
		
	\end{figure*}
	In Fig. \ref{fig:Val100}, for $T = \SI{100}{\micro\second}$, we compare the rate-power regions of the \gls*{swipt} system obtained for Schemes I-III and the baseline scheme from \cite{Morsi2019} in the \gls*{spr} and \gls*{mpr} regimes. 
	To this end, for all considered schemes, we show the expected mutual information $\overline{I}\big(p_{r_x}^*\big)$ and the average harvested power $\overline{P}\big(p_{r_x}^*\big)$ obtained with the circuit simulator ADS for the respective optimal input distribution $p_{r_x}^*$. 
	We observe that in the \gls*{spr} regime, all four schemes yield a similar performance since, in this case, the \gls*{eh} operates in the linear regime and the rectenna memory can be neglected because of the low input power and the large symbol duration, cf. Section \ref{SectionInfiniteDuration}, respectively.
	However, in the \gls*{mpr} regime, the saturation of the \gls*{eh} has an impact on performance. 
	Thus, in this regime, the baseline scheme, which is based on an analytical model for the \gls*{eh} circuit, has a worse performance than the other schemes, which employ a learning based model for the \gls*{eh} circuit and are able to better capture the impact of impedance mismatch and \gls*{eh} saturation.	
	
	In Fig. \ref{fig:ValMP}, for the \gls*{mpr} regime, we compare the rate-power regions of the considered schemes for symbol durations $T~=~\SI{100}{\micro\second}$ and $T = \SI{10}{\micro\second}$, respectively.	
	We observe that for all considered schemes, a shorter symbol duration generally leads to lower harvested powers.	
	Furthermore, unlike for $T = \SI{100}{\micro\second}$, for $T = \SI{10}{\micro\second}$, there is a significant performance gap between Scheme I, for which the \gls*{eh} state is known at \gls*{tx} and \gls*{ir}, and the other schemes, for which the \gls*{eh} state is unknown at both devices.	 
	This is expected, since, for short $T$, the memory introduced by the \gls*{eh} is significant, which is exploited in Scheme I by finding the optimal input distribution for each \gls*{eh} state. 
	In contrast, the other schemes have to find a compromise input distribution which yields a good performance for all the \gls*{eh} states.		
	Additionally, we observe a larger performance gap between Schemes II and III for the shorter symbol duration since the \gls*{eh} memory is completely neglected in Scheme III. 
	Finally, we note that for a given symbol duration $T$ and a given power regime, the input distributions and, hence, the harvested powers for all considered schemes are nearly identical for large $\overline{I}\big(p_{r_x}^*\big)$ since, in this case, the solution of (\ref{MajorOptProblem}) is mainly determined by the feasible set specified by (\ref{MajorOptProblem_Constr1})-(\ref{MajorOptProblem_Constr5}).	
	\subsection{Impact of Impedance Mismatch}
	\label{SectionImpactMismatch}	
	\begin{figure*}[!t]	
		\centering
		\includegraphics[width=0.98\textwidth]{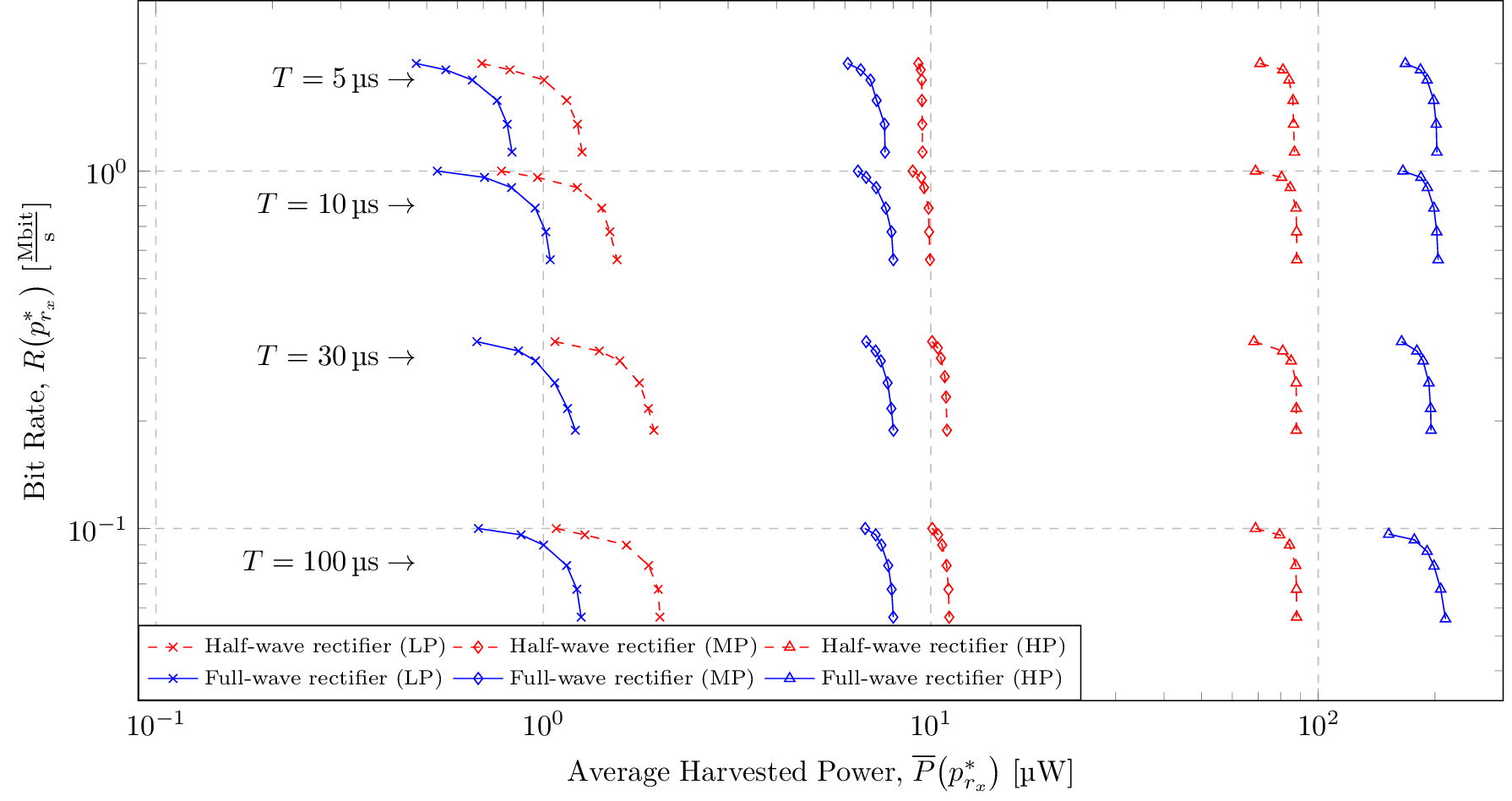} 		
		\caption{Rate-power region for Scheme II, different symbol durations $T$, input power regimes, and rectifier circuits.}
		\label{MI_AHP_Fig} 
		\centering
	\end{figure*}		
	In this section, we study the impact of impedance mismatch between antenna and rectifier on the performance of the \gls*{swipt} system. 
	To this end, in Fig. \ref{Mismatch_Fig}, we show the rate-power regions for Schemes I and II for the \gls*{spr}, \gls*{mpr}, and \gls*{lpr} regimes. 
	For the results shown in Fig. \ref{Mismatch_Fig}, we adopt a symbol duration of $T = \SI{10}{\micro\second}$ and the MCs were tuned for two different input power levels, namely, $\SI{-13}{\dBm}$ and $\SI{0}{\dBm}$.  
	We show in Fig. \ref{fig:MatchingHWR} and Fig. \ref{fig:MatchingFWR} the rate-power regions obtained for half-wave and full-wave rectifiers, respectively.	
	We observe from Fig. \ref{Mismatch_Fig} that as expected, a larger input power generally leads to a higher average harvested power $\overline{P}$. 
	Moreover, we note that in the \gls*{spr} and \gls*{mpr} regimes, for both rectifier circuits, the impedance mismatch caused by employing an MC fine-tuned for a relatively high input power level of $\SI{0}{\dBm}$ yields a significant performance loss compared to an MC designed for a lower input power level of $\SI{-13}{\dBm}$. 
	In contrast, in the \gls*{lpr} regime, the MC tuned for $\SI{0}{\dBm}$ outperforms the MC tuned for $\SI{-13}{\dBm}$ for both rectifier circuits. 
	However, the performance difference is much larger for the full-wave rectifier than for the half-wave rectifier. 
	In fact, in the \gls*{lpr} regime, the \gls*{eh} employing the half-wave rectifier is almost always driven into saturation which limits the amount of harvested power.	
	Additionally, we note that similar to Fig. \ref{Validation_Fig}, for a given rectifier circuit at the \gls*{eh} and a given input power regime, exploiting the knowledge of the \gls*{eh} state, i.e., Scheme I, typically yields a performance gain.
	\subsection{Influence of Symbol Duration and Input Power Regime}	
	\setcounter{equation}{22}	
	\begin{figure*}[!ht]
		\normalsize
		\begin{equation}
		\begin{aligned}
		p_{r_y}\big(r_y; p_{r_x}\big) &= \int_{0}^{r_x^\text{max}} \int_{-\pi}^{\pi} \int_{-\pi}^{\pi} p_{r_y, \phi_y \mid r_x, \phi_x} (r_y, \phi_y \mid r_x, \phi_x)  \frac{1}{2 \pi}p_{r_x}(r_x)  d\phi_x \, d\phi_y \, dr_x  \\[-0.2ex]
		&= \frac{1}{\sigma_n^2} \int_{0}^{r_x^\text{max}} r_y e^{-\frac{r_y^2 + r_x^2 |h_I|^2}{2 \sigma_n^2}} I_0 \bigg( \frac{r_y \, r_x \, |h_I|}{\sigma_n^2} \bigg) p_{r_x}(r_x) dr_x.
		\end{aligned}
		\label{Appendix_OutputDistributionEqn}
		\end{equation}		
		\begin{equation}
		\begin{aligned}
		&I\big( p^i_{r_x, \phi_x} \big) = I\big( p^i_{r_x}, p^i_{\phi_x \mid r_x} \big) = H_{r_y,\phi_y}\big( p^i_{r_x, \phi_x} \big)  + \int_0^{\infty} p_{r_y}\big(r_y; p^i_{r_x, \phi_x}\big) \log_2 (r_y) dr_y - H_n  \\[-0.8ex]
		&\leq I\big( p^i_{r_x}, p^i_{\phi_x \mid r_x} = p^{\text{uni}}_{\phi_x} \big) = H_{r_y}\big( p^i_{r_x} \big) + \int_0^{\infty} p_{r_y}\big(r_y; p^i_{r_x}\big) \log_2 (r_y) dr_y + \log_2 (2 \pi) - H_n,
		\end{aligned}
		\label{AppendixA_Constraint}
		\end{equation}				
		\hrulefill
		\vspace*{4pt}
	\end{figure*}	
	\setcounter{equation}{20}
	In this section, we study the impact of symbol duration $T$ and the input power regime on the performance of Scheme II, i.e., when the \gls*{eh} state is not known at \gls*{tx} and \gls*{ir}. 
	We assume that the \gls*{eh} is equipped with a half- or a full-wave rectifier and the impedance MC is designed for the corresponding input power levels, i.e., the MCs are tuned for $\SI{-13}{\dBm}$ in the \gls*{spr} and \gls*{mpr} regimes and for $\SI{0}{\dBm}$ in the \gls*{lpr} regime, cf. Fig. \ref{Mismatch_Fig}.		
	In Fig. \ref{Distributions_Fig}, we show the optimal distributions for Scheme II in the \gls*{spr} and \gls*{mpr} regimes for symbol durations $T = \SI{100}{\micro\second}$ and $T = \SI{10}{\micro\second}$. 
	The required mutual information was set to $I_{\text{req}} = \SI{6.5}{\frac{\bit}{symbol}}$.	
	We observe that for both the half-wave rectifier (Fig. \ref{fig::DistrHWR}) and the full-wave rectifier (Fig. \ref{fig::DistrFWR}), the optimal input distribution is practically independent of symbol duration $T$ in the \gls*{spr} regime, where it is optimal to allocate a probability of $0.12$ and $0.14$ to symbols having the maximum amplitude $r_x^\text{max}$, respectively, as even for this large amplitude, the \gls*{eh} circuit is not in saturation. 
	
	However, in the \gls*{mpr} regime, for both symbol durations, the symbol amplitudes for the half-wave rectifier are limited to values smaller than $r_x^\text{max}$ to avoid that the \gls*{eh} circuit is driven into saturation.	
	Furthermore, the optimal input distribution depends on the value of the symbol duration.
	For short symbol durations, the capacitor $C_L$ in the \gls*{eh} circuit cannot be fully charged within one symbol interval, and hence, larger symbol amplitudes can be afforded without driving the \gls*{eh} into saturation.			
	Similarly, for the full-wave rectifier, in the \gls*{mpr} regime, the optimal input distribution favors smaller amplitudes for $T = \SI{100}{\micro\second}$. 
	However, the symbol amplitudes for the full-wave rectifier in the \gls*{mpr} regime tend to be larger than those for the half-wave rectifier.		
	This is due to the larger breakdown voltage of the former, where two identical diodes are connected in series, compared to the latter, which has only a single diode.
	Hence, in the \gls*{mpr} regime, the optimal input distribution depends on the symbol duration and the rectifier circuit.	
		
	In Fig.~\ref{MI_AHP_Fig}, we show the boundaries of the rate-power region for Scheme II. 
	Here, in order to be able to illustrate the impact of the symbol duration on the data rate, we show the bit rate, $R\big(p_{r_x}^*\big) = {I\big(p_{r_x}^*\big)}/{T}$, as a function of the harvested power $\overline{P}\big(p_{r_x}^*\big)$.
	For the results shown in Fig.~\ref{MI_AHP_Fig}, we adopted Rayleigh fading for the \gls*{ir} channel, whereas, for the \gls*{eh} channel, we assumed a line of sight and, hence, Rician fading with Rician factor 1. The simulation results were averaged over 1000 channel realizations.	
	In Fig.~\ref{MI_AHP_Fig}, the rate-power regions for different symbol durations and for the \gls*{spr}, \gls*{mpr}, and \gls*{lpr} regimes are depicted. 	
	First, similar to Fig. \ref{Mismatch_Fig}, we observe that for both considered \gls*{eh} circuits and all considered symbol durations, $T$, higher input power levels lead to higher average harvested powers. 
	Furthermore, for all considered symbol durations, $T$, the half-wave rectifier yields a better performance compared to the full-wave rectifier in the \gls*{spr} and \gls*{mpr} regimes due to the smaller number of lossy non-linear diodes.
	However, in the \gls*{lpr} regime, the full-wave rectifier performs better since the two diodes connected in series lead to a higher power saturation level. 	
	Furthermore, for all input power regimes and for both \gls*{eh} circuits, decreasing the symbol duration generally leads to an increase of the bit rate at the \gls*{ir} and a reduction of the average power that can be harvested by the \gls*{eh}. 
	In particular, for the half-wave rectifier in the \gls*{mpr} regime, average harvested power values larger than $\SI{10}{\micro\watt}$ can be achieved only if the symbol duration exceeds $T = \SI{10}{\micro\second}$. 
	Hence, Fig.~\ref{MI_AHP_Fig} reveals that the rate-power region of the considered SWIPT system depends on the symbol duration, the input power level at the \gls*{eh}, and the type of \gls*{eh} circuit since the memory and non-linearity of the rectenna have a significant impact on the amount of harvested power. 		
	\section{Conclusion}
	\label{SectionConclusion}
	In this paper, we considered \gls*{swipt} systems that employ \gls*{eh}s with practical non-linear rectifier circuits with memory and impedance mismatch. 
	We modeled the memory of the \gls*{eh} by an \gls*{mdp} and used \gls*{dnn}s to model the non-linear effects of the \gls*{eh} circuit.
	For optimization of the input symbol distribution, we considered the cases where \gls*{tx} and \gls*{ir} know and do not know the \gls*{eh} state. 
	We showed that for the optimal input symbol distribution, the phase of the transmit signal is independent of the signal amplitude and uniformly distributed. 
	Furthermore, for the case where \gls*{tx} and \gls*{ir} know the instantaneous \gls*{eh} state, we formulated a convex optimization problem to determine the boundary of the rate-power region.
	Then, for the case where the \gls*{eh} state is not known at \gls*{tx} and \gls*{ir}, we showed that the corresponding optimization problem is non-convex and proposed an iterative algorithm based on alternating optimization to obtain a limit point.
	In our simulation results, we considered \gls*{eh}s with half-wave and full-wave rectifier circuits. 
	We validated our model by comparing it with a baseline scheme and studied the impact of the symbol duration, the \gls*{eh} input power level, the impedance mismatch between antenna and rectifier, and the type of \gls*{eh} circuit.
	We observed that in the \gls*{spr} and \gls*{mpr} regimes, the half-wave rectifier circuit yields a larger average harvested power than the full-wave rectifier, whereas, in the \gls*{lpr} regime, the latter circuit significantly outperforms the former.
	Additionally, our results showed that for both rectifier circuits and all considered input power regimes, a shorter symbol duration leads to a higher bit rate at the expense of a decrease of the average harvested power.	
	
	{The extension of the proposed \gls*{mdp} framework to \gls*{swipt} systems with arbitrary transmit pulse shapes and \gls*{eh} circuits comprising multiple memory elements as part of the \gls*{lp} or a battery storage are interesting directions for future research.}
	\appendices
	\section{Proof of Proposition \ref{Theorem1}} 
	\label{AppendixA}		
	The following proof follows \cite[Section II.B]{Shamai1995}.
	First, let us observe that for any concave function $f\big(\cdot; \cdot \big)$, and any $F \in \mathbb{R}$,	
	\begin{align}
	&\text{if} \; F \leq f\big(p_1; p_2 \big) \leq \;  f\big( p_1; p_2 = p'_2 \big), \; \forall p_1, p_2, \nonumber \\ &\begin{aligned}\text{then} \; \{p_1 \mid f\big(p_1; p_2 \big) &\geq F \} \\ 
	&\subseteq \{p_1 \mid f\big( p_1; p_2 = p'_2 \big) \geq F \},\end{aligned}
	\label{AppendixA_Observation}
	\end{align}		
	\noindent where $p_1 \in \mathbb{P}_1$, $p_2, p'_2 \in \mathbb{P}_2$, and $\mathbb{P}_1, \mathbb{P}_2$ are some sets of functions $\mathbb{R} \mapsto \mathbb{R}$.
	
	Then, we note that the joint \gls*{pdf} of \gls*{rv}s $r_y$ and $\phi_y$ conditioned on $r_x$ and $\phi_x$ for the considered \gls*{awgn} channel can be expressed as \cite[Eq. (10)]{Shamai1995} 
	\begin{equation}
	\begin{aligned}
		p_{r_y, \phi_y \mid r_x, \phi_x} (r_y, &\phi_y \mid r_x, \phi_x) = \frac{r_y}{2 \pi \sigma_n^2}\\
		&\times e^{-\frac{r_y^2 + r_x^2 |h_I|^2 - 2 r_x |h_I| \, r_y \cos(\phi_y - \phi_x - \phi_I)}{2 \sigma_n^2}  }.
	\end{aligned}
	\end{equation} 			
	
	Therefore, if the pdf of phase $\phi_x \in (-\pi, \pi]$ is uniform and independent of $r_x$, then taking constraint (\ref{MajorOptProblem_Constr3}) into account, we obtain the marginal pdf $p_{r_y}$ as a function of the input \gls*{pdf} $p_{r_x}$ as in (\ref{Appendix_OutputDistributionEqn}), shown on top of this page.
	\setcounter{equation}{24}	
	Thus, if amplitude $r_x$ and phase $\phi_x$ are statistically independent and $\phi_x$ is uniformly distributed, then amplitude $r_y$ and phase $\phi_y$ of the received signal are also mutually statistically independent. 
	Moreover, in this case, the phase of the received signal is uniformly distributed and the \gls*{pdf} of the amplitude of the received signal, $r_y$, is given by (\ref{Appendix_OutputDistributionEqn}).
	
	Then, we note that the joint differential entropy of \gls*{rv}s $r_y$ and $\phi_y$ is always bounded as \cite{Shamai1995} $H_{r_y, \phi_y}~\leq~H_{r_y}~+~\log_2 (2\pi)$, where $H_{r_y}$ denotes the entropy of $r_y$. 
	Moreover, the relation is satisfied with equality if $\phi_y$ is uniformly distributed and independent of $r_y$, or equivalently, if $\phi_x$ is uniformly distributed and independent of $r_x$.	
	Hence, the mutual information achieved by input pdf $p^i_{r_x, \phi_{x}}$ is given by (\ref{AppendixA_Constraint}), shown on top of this page, where $p^{\text{uni}}_{\phi_x}$ denotes the uniform pdf of $\phi_x$. In (\ref{AppendixA_Constraint}), equality holds if $r_x$ and $\phi_x$ are independent, $\phi_x$ is uniformly distributed and, hence, is independent of \gls*{eh} state $\xi$. 
	
	Furthermore, we note that the average mutual information $\overline{I}$ in (\ref{MajorOptProblem_Constr1}) is defined as a weighted sum of the individual values $I\big( p^i_{r_x, \phi_x} \big)$, i.e., $\overline{I}\big( \boldsymbol{\pi}, \mathcal{P}_{\phi_x \mid r_x} \big) = \sum_{i=1}^{S_\Xi} {\gamma}_i I\big( p^i_{r_x, \phi_x} \big)$. Hence, we can rewrite constraint (\ref{MajorOptProblem_Constr1}) as follows	
	\begin{equation}
	\begin{aligned}
		I_{\text{req}} &\leq \overline{I}\big( \boldsymbol{\pi}, \mathcal{P}_{\phi_x \mid r_x} \big) = \sum_{i=1}^{S_\Xi} {\gamma}_i I\big( p^i_{r_x, \phi_x} \big) \\[-1.8ex]
		&\leq \sum_{i=1}^{S_\Xi} {\gamma}_i I\big( p^i_{r_x}, p^i_{\phi_x \mid r_x} = p^{\text{uni}}_{\phi_x} \big) = \overline{I}( \boldsymbol{\pi}).
	\end{aligned}
	\label{AppendixA_GeneralInequality}	
	\end{equation}	
	Since $I\big( \cdot, \cdot \big)$ and, hence, $\overline{I}\big( \cdot, \cdot \big)$ in (\ref{AppendixA_GeneralInequality}) are concave, then, from~(\ref{AppendixA_Observation}),	
	\begin{equation}
	\begin{aligned}
	\{\boldsymbol{\pi} \mid  \overline{I}\big( \boldsymbol{\pi}, &\mathcal{P}_{\phi_x \mid r_x} \big) \geq	I_{\text{req}} \} \\
	& \subseteq \{\boldsymbol{\pi} \mid \overline{I}\big( \boldsymbol{\pi}; p^i_{\phi_x \mid r_x} = p^{\text{uni}}_{\phi_x} \big) \geq I_{\text{req}} \}.
	\end{aligned}
	\end{equation}	
	Thus, independently distributed \gls*{rv}s $r_y$ and $\phi_y$ with uniformly distributed $\phi_y$ lead to the largest set of pdfs $\boldsymbol{\pi}(r_x)$ with $ \overline{I}\big( \boldsymbol{\pi}, \mathcal{P}_{\phi_x \mid r_x} \big)\geq I_{\text{req}}$. 
	Since the objective function and the other constraints in (\ref{MajorOptProblem}) do not depend on $\mathcal{P}_{\phi_x \mid r_x}$, this condition also leads to the largest feasible set of pdfs $\boldsymbol{\pi}(r_x)$ for optimization problem (\ref{MajorOptProblem}). 
	Therefore, for the optimal solution of (\ref{MajorOptProblem}), the phase $\phi_x$ is uniformly distributed and statistically independent of the state $\xi$ and the amplitude $r_x$. 
	Moreover, the marginal \gls*{pdf} of the output symbol amplitudes, $p_{r_y}\big(r_y; p_{r_x}\big)$, is given by (\ref{Appendix_OutputDistributionEqn}), whereas the mutual information in (\ref{AppendixA_Constraint}) can be simplified to the expression in (\ref{ProposedMutualInf})
	This completes the proof. \\	
	\bibliographystyle{IEEEtran}
	\bibliography{JourPaper1_IEEE}	
	\begin{IEEEbiography}[{\includegraphics[width=1in,height=1.25in,clip,keepaspectratio]{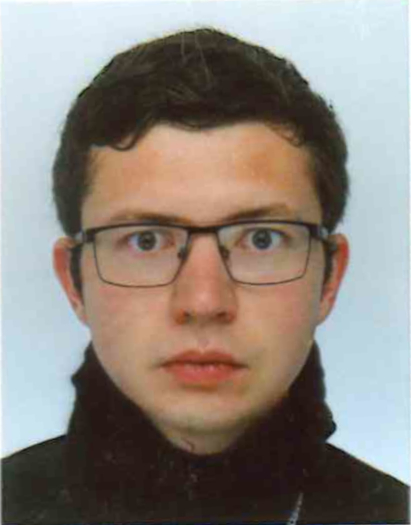}}]
		{Nikita Shanin} (S'18) received the Diplom degree from the Bauman Moscow State Technical University (BMSTU), Moscow, Russia in 2018. He is currently working towards a Ph.D degree at the Institute for Digital Communications of the Friedrich-Alexander Universit\"{a}t (FAU), Erlangen-N\"{u}rnberg, Germany. His research interests fall into the broad areas of signal processing and wireless communications, including wireless information and power transfer.
	\end{IEEEbiography}
	
	\begin{IEEEbiography}[{\includegraphics[width=1in,height=1.25in,clip,keepaspectratio]{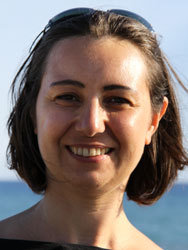}}]{Laura Cottatellucci} (S'01-M'07) received the Master degree from La Sapienza University, Rome, Italy, the Ph.D. degree from the Technical University of Vienna, Austria, in 2006, and the Habilitation degree from the University of Nice-Sophia Antipolis, France. From December 2017, she is Professor for Digital Communications at the Institute of Digital Communications of Friedrich Alexander Universit\"{a}t (FAU) of Erlangen-N\"{u}rnberg (Germany). She worked in Telecom Italia (1995-2000) as responsible of industrial projects and as a Senior Research in Forschungszentrum Telekommunikation Wien Austria (Apr. 2000-Sep. 2005). She was a Research Fellow in INRIA, France, (Oct.-Dec. 2005) and at the University of South Australia (2006). She was Assistant Professor (Dec. 2006-Nov. 2017) and subsequently Adjunct Professor (Mar. 2006-Aug. 2017) in EURECOM, France. Her research interests lie in the field of communications theory and signal processing for wireless communications, satellite and complex networks. She served as associate editor of the IEEE Transactions on Communications and the IEEE Transactions on Signal Processing (Feb 2016-2020). She is an elected member of the IEEE Technical Committee on Signal Processing for Communications and Networking since 2017.
	\end{IEEEbiography}

	\begin{IEEEbiography}[{\includegraphics[width=1.4in,height=1.3in,trim=0in 1.5in 0in 0.5in,clip,keepaspectratio]{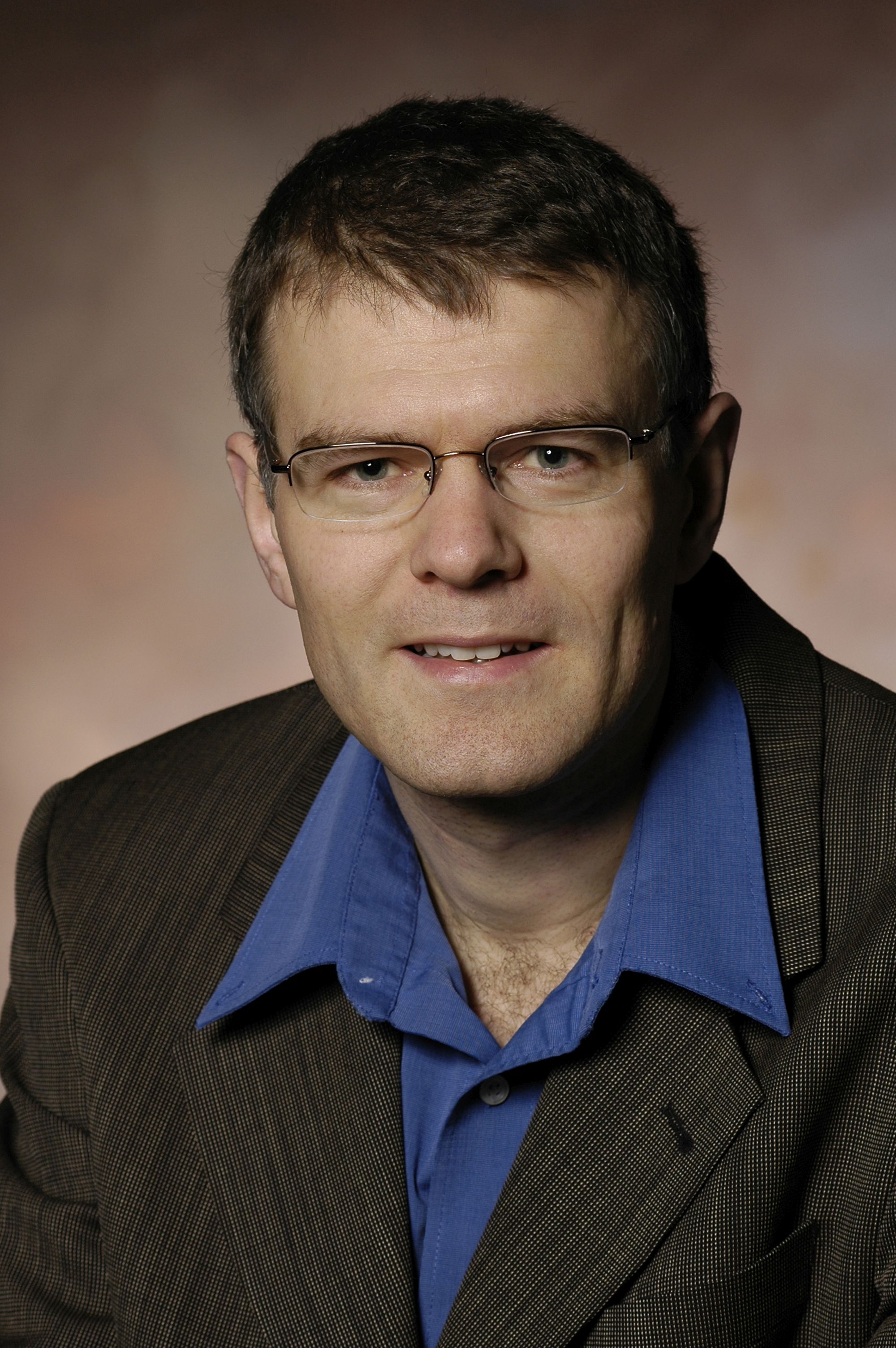}}]{Robert Schober} (S'98, M'01, SM'08, F'10) received the Diplom (Univ.) and the Ph.D. degrees in electrical engineering from Friedrich-Alexander	University of Erlangen-Nuremberg (FAU), Germany, in 1997 and 2000, respectively. From 2002 to 2011, he was a Professor and Canada Research Chair at the University of British Columbia (UBC), Vancouver, Canada. Since January 2012 he is an Alexander von Humboldt Professor and the Chair for Digital Communication at FAU. His research interests fall into the broad areas of Communication Theory, Wireless Communications, and Statistical Signal Processing.
		
	Robert received several awards for his work including the 2002 Heinz Maier­ Leibnitz Award of the German Science Foundation (DFG), the 2004 Innovations Award of the Vodafone Foundation for Research in Mobile Communications, a 2006 UBC Killam Research Prize, a 2007 Wilhelm Friedrich Bessel Research Award of the Alexander von Humboldt Foundation, the 2008 Charles McDowell Award for Excellence in Research from UBC, a 2011 Alexander von Humboldt Professorship, a 2012 NSERC E.W.R. Stacie Fellowship, and a 2017 Wireless Communications Recognition Award by the IEEE Wireless Communications Technical Committee. Since 2017, he has been listed as a Highly Cited Researcher by the Web of Science. Robert is a Fellow of the	Canadian Academy of Engineering and a Fellow of the Engineering Institute of Canada. From 2012 to 2015, he served as Editor-in-Chief of the IEEE	Transactions on Communications. Currently, he serves as Member of the Editorial Board of the Proceedings of the IEEE and as VP Publications for the IEEE Communication Society (ComSoc).
	\end{IEEEbiography}
	
\end{document}